\documentclass[twocolumn,superscriptaddress]{revtex4}

\usepackage{graphicx}                   

\usepackage{threeparttable} 
\usepackage{array}

\usepackage{color}                       

\usepackage{float}                       
\usepackage[hidelinks]{hyperref}         
\usepackage{wrapfig}                     

\usepackage[parfill]{parskip}            

\usepackage{amstext,amsmath,amssymb,amsfonts,braket}                 
\usepackage{siunitx}

\usepackage{xspace}

\usepackage[normalem]{ ulem }
\usepackage{soul}

\newcommand{\beq}{\begin{equation}}
\newcommand{\eeq}{\end{equation}}

\newcommand{\beqq}{\begin{equation*}}
\newcommand{\eeqq}{\end{equation*}}

\newcommand{\bcen}{\begin{center}}
\newcommand{\ecen}{\end{center}}

\newcommand{\tsp}{\textsuperscript}
\newcommand{\tsb}{\textsubscript}
\newcommand{\tsc}{\textsc}

\newcommand{\dch}{K$^{-2}$\xspace}
\newcommand{\Chi}{\mathcal{X}\xspace}
\newcommand{\COO}{CO\textsubscript{2}\xspace}

\begin{document}

\title{Specific chemical bond relaxation unravelled by analysis of shake-up satellites in the oxygen single site double core hole spectrum of \COO}

\author{Anthony Fert\'e}\email{anthony.ferte@sorbonne-universite.fr}
\affiliation{Laboratoire de Chimie Physique Mati\`ere et Rayonnement (LCPMR), Sorbonne Universit\'e and CNRS, F-75005 Paris, France}
\author{Francis Penent}
\affiliation{Laboratoire de Chimie Physique Mati\`ere et Rayonnement (LCPMR), Sorbonne Universit\'e and CNRS, F-75005 Paris, France}
\author{J\'er\^ome Palaudoux}
\affiliation{Laboratoire de Chimie Physique Mati\`ere et Rayonnement (LCPMR), Sorbonne Universit\'e and CNRS, F-75005 Paris, France}
\author{Hiroshi Iwayama}
\affiliation{UVSOR Facility, Institute for Molecular Science, Okazaki 444-8585, Japan}
\author{Eiji Shigemasa}
\affiliation{UVSOR Facility, Institute for Molecular Science, Okazaki 444-8585, Japan}
\author{Yasumasa~Hikosaka}
\affiliation{Institute of Liberal Arts and Sciences, University of Toyama, Toyama 930-0194, Japan}
\author{Kouichi Soejima}
\affiliation{Department of Environmental Science, Niigata University, Niigata, 950-2181, Japan}
\author{Pascal Lablanquie}
\affiliation{Laboratoire de Chimie Physique Mati\`ere et Rayonnement (LCPMR), Sorbonne Universit\'e and CNRS, F-75005 Paris, France}
\author{Richard Ta\"ieb}
\affiliation{Laboratoire de Chimie Physique Mati\`ere et Rayonnement (LCPMR), Sorbonne Universit\'e and CNRS, F-75005 Paris, France}
\author{St\'ephane Carniato}\email{stephane.carniato@upmc.fr}
\affiliation{Laboratoire de Chimie Physique Mati\`ere et Rayonnement (LCPMR), Sorbonne Universit\'e and CNRS, F-75005 Paris, France}

\date{\today}

\begin{abstract}
We developed recently [A. Fert\'e, \textit{et al.}, J. Phys. Chem. Lett. 11, 4359 (2020)] a method to compute single site double core hole (ssDCH or \dch) spectra. We refer to that method as NOTA+CIPSI. In the present paper this method is applied to the O \dch spectrum of the \COO molecule, and we use this as an example to discuss in detail its convergence properties. Using this approach, a theoretical spectra in excellent agreement with the experimental one is obtained. Thanks to a thorough interpretation of the shake-up states responsible for the main satellite peaks and with the help of a comparison with the O \dch spectrum of CO, we can highlight the clear signature of the two non equivalent carbon oxygen bonds in the oxygen ssDCH \COO dication.
\end{abstract}

\maketitle

\section{Introduction}

In the late eighties, Cederbaum \textit{et al.\,} conducted theoretical investigations \cite{CedTarSch-JCP-86,CedTarSch-JCP-87,Ced-PRA-87} showing that a new spectroscopy, based on double core hole (DCH) states, could be more sensitive than the common single core hole x-ray photoelectron spectroscopy (SCH-XPS) \cite{NorSokSie-PR-57,SokNorSie-PR-58,SieNorFah-ESCA-67} that remains, nowadays, one of the most applied technique for chemical analysis.
In addition to the chemical specificity and to the sensitivity to the chemical environment that it shares with SCH-XPS, DCH spectroscopy could display an enhanced sensitivity to the bond length \cite{TarSgaCed-JCP-87} and would also open new possibilities when two core holes are located on different atoms. However, it was only during the past decade that the experimental observation of such doubly core ionized molecular systems was achieved, either arising from single x-ray photon absorption at synchrotron facilities \cite{LabPenHik-JPB-16}, or through the sequential absorption of two x-ray photons that became possible with X-FEL (X-ray Free Electron Laser) sources~\cite{BerFan-JESRP-15}.

When the two ejected core electrons originate from the same atom, leading to the formation of a single site double core hole (ssDCH or \dch) states, a strong increase of the shake-up satellite peaks is observed. Such satellites result from the excitation of an outer electron in conjunction with the core photoionization \cite{Blo-PR-35,Abe-PR-67,MelPer-PRA-71,AarGueHil-JCSFT-72,BriBak-JESRP-75,RosWerGug-PRL-71,MarShi-JCP-76}. The multielectronic character of the shake-up process implies that these states are associated with important electronic relaxation and correlation effects. The fundamental nature of these effects justifies our interest in studying shake-up satellite states.

 Moreover, these shake-up states are of critical importance when computing ssDCH spectra as individual shake-up lines are of comparable intensity with the main \dch peak. 
 In comparison, individual shake-up typically represents only few percents of the main line in a standard SCH-XPS spectrum (see \textit{e.g.} Refs. \cite{EhaTanUed-JCP-06,SanEhaUed-CPL-06,CarSelPen-PRA-16} and supplementary material where we report the experimental intensity ratio between K\tsp{$-1$} main and shake-up lines for \COO which were not given in Ref. \cite{CarSelPen-PRA-16}\,).
Therefore, these states must be accurately described to obtain results that compare well with the experimental data. However, their accurate computation requires a method that correctly accounts for the electronic relaxation in the final ssDCH as well as for specific correlation effects. In addition, one needs to take into account a large number of excited states in order to compute the complete spectrum. This makes the computation of \dch shake-up spectra challenging and pushes quantum chemistry methods to their limits.

For conventional SCH-XPS, many different computational strategies have been used. This includes wave function theory approaches such as unrestricted Hartree-Fock (UHF) \cite{AarBarGue-MP-73}, static exchange (STEX) \cite{CarAgrNor-CPL-88}, configuration interaction (CI) \cite{GueHilWoo-PRSA-73,Woo-CP-74,MarShi-PRA-76}, selected CI \cite{GueRodHil-JCP-77}, multi configuration self consistent field (MCSCF) \cite{Bas-JESRP-74,NorNilAgr-JESRP-91,CarDufRoc-JESRP-94}, Green's function algebraic-diagrammatic construction (ADC) \cite{AngWalSch-JCP-87} and coupled cluster (CC) \cite{CorKoc-JCP-15}. On the other hand, strategies based on density functional theory (DFT), such as time dependent density functional theory (TDDFT) \cite{BreLuoCar-PRB-04}, have also been considered and sub-eV error was recently achieved on the determination of core ionization potential using a restricted open shell Kohn-Sham approach coupled with the square gradient minimization method~\cite{HaiGor-JPCL-20}.

~

However, up to now, only few of these methods have been transposed to ssDCH spectroscopy and most of the existing work only focused on computing  double core ionization potentials (DCIP) \cite{TasEhaCed-JCP-10,TasEhaUed-CPL-10,LabCarIto-PRL-11,TakTasEhaUed-JPCA-11,UedTak-JESRP-12,NakCarIto-PRL-13,TakKryUed-JESRP-15,LeeSmaGor-JCP-19}. For the computation of complete ssDCH spectra (including transition amplitudes and shake-up peaks), second order ADC(2) \cite{SanKryCed-PRL-09} and truncated CI \cite{TasUedEha-CPL-11,CarSelFerBerWuoNakHikItoZitBucAndPalPenLab-JCP-19} methods have been used. Both approaches showed some limitations and the final result often failed to accurately reproduce the position and the intensity of all  peaks. In particular, ADC(2) method suffers from its inability to fully account for the relaxation induced by the double core ionization \cite{OhrCedTar-PRA-91,KrySanCed-JCP-11}, while arbitrarily truncated CI method may experience difficulties in balancing the description of all the important \dch states.

In our previous paper \cite{FerPalPenIwaShiHikSoeItoLabTaiCar-JPCL-20}, an original computation method was developed to answer the theoretical questions raised by the computation of {ssDCH} spectra. This method called non orthogonal computation of transition amplitude with CIPSI wave functions (NOTA+CIPSI) combines a selected configuration interaction method (CIPSI) and the use of two sets of non mutually orthogonal molecular orbitals to improve the convergence of transition probabilities. It was applied to compute the O \dch spectrum of CO and we showed that it reproduces with a high level of accuracy the experimental result both in terms of binding energies and of relative cross sections and thus represents a major improvement with respect to previously available results. We also highlighted that the high intensity associated with the main satellites in a ssDCH spectrum can be understood as the consequence of a compensation between the double core hole induced electronic relaxation and the shake-up induced electronic reorganization.

In section~\ref{theory}, we recall the underlying theory and the main characteristics of the NOTA+CIPSI method. 
In section~\ref{comput_details}, we give computational details regarding the presented calculations. 
Convergence and reliability of the present results are assessed in \ref{convergence_section}. 
We thoroghly investigate the O ssDCH spectrum of CO\tsb2 in subsection~\ref{CO2}. This includes the examination of vibronic and geometrical relaxation effects and a thorough interpretation of the shake-up satellites.
Then, we highlight the interesting points of comparison between the CO and \COO \dch spectra in~\ref{CO_vs_CO2}. 
Finally, we conclude by a summary of our results in section~\ref{conclusion}.
 
Atomic units are used throughout this paper if no specification is given.

\section{Theory}
\label{theory}

\subsection{Single site double core ionization cross section in the dipole approximation}
Within the framework of the dipole approximation and in the length gauge, the general expression of the differential cross section for a single site double core ionization is 
\begin{align}
\label{cross_dip}
\frac{\mathrm{d}^6\sigma_{\tsc{i} \rightarrow \tsc{f}}}{\mathrm{d}\mathbf{k}_1\mathrm{d}\mathbf{k}_2} = 4 \pi \alpha \omega &\times \left|\Braket{\Psi_\textsc{f}(N)| \sum_{i=1}^N \mathbf{r}_i | \Psi_\textsc{i}(N)}\right|^2 \nonumber\\ 
&\times \mathcal{L}(\omega-(E_\tsc{f} + \epsilon);\Gamma_\tsc{f})
\end{align}
where $\Psi_\textsc{i}(N)$ and $\Psi_{\tsc{f}}(N)$ are $N$ electron wave functions respectively describing the initial neutral state and a final ssDCH state including the two ejected electrons of asymptotic momenta $\mathbf{k}_1$ and $\mathbf{k}_2$ respectively. Here, $\omega$ is the energy of the photon, $E_\tsc{f}$ is the binding energy of the final \dch state, $\epsilon$ is the total kinetic energy carried out by the two photoelectrons and, finally, $\mathcal{L}(\omega-(E_\tsc{f} + \epsilon);\Gamma_\tsc{f})$ is the Lorentzian function accounting for the lifetime $\Gamma_\tsc{f}$ of the final dicationic state. The form of the final ssDCH state wave function is taken as
\beq
\Psi_\textsc{f}(N) = \hat{\mathcal{A}}(N) \Chi_{\mathbf{k}_1,\mathbf{k}_2}(1,2) \Psi_\textsc{f}^{\tsc{k-2}}(N-2)
\eeq
where $\hat{\mathcal{A}}(N)$ is the $N$ electron antisymmetrizer operator, $\Chi_{\mathbf{k}_1,\mathbf{k}_2}(1,2)$ is a double continuum wave function describing two free electrons and $\Psi_\textsc{f}^{\tsc{k-2}}(N-2)$ is the ($N-2$) electron wave function describing the remaining electrons. Since the correlation between the two outgoing electrons and the bound ones can be considered as negligible, $\Psi_\textsc{f}^{\tsc{k-2}}$ is assimilated to the wave function of the isolated \mbox{ssDCH}. Moreover, as the $\Psi_\textsc{f}(N)$ wave function is normalized, a condition reminiscent of the strong orthogonality ansatz need to be fulfilled~\cite{CedDomSch-ACP-86}, further bolstering the independence between the ``bound" wave function and two-electron continuum part.

The intensity of a peak associated with the formation of one specific ssDCH state by absorption of a single x-ray photon is proportional to the square of the dipole transition moment between the initial and the final wave functions. However, considering that (\textit{i}) the two electron dipole term arising from the two outgoing electrons is mostly independent of the final ssDCH state (because of the weak correlation between the ejected electrons and the bound ones), (\textit{ii}) the density of continuum states is almost constant for photon energy far above the DCIP, the transition moment can be simplified as
\beq
\label{overlap}
T_{\textsc{i} \rightarrow \textsc{f},n} \propto \Braket{\Psi_{\textsc{f},n}^\textsc{\tsc{k-2}}(N-2)|\hat{a}_{1\mathrm{s}}^\alpha \hat{a}_{1\mathrm{s}}^\beta|  \Psi_\tsc{i}(N)}.
\eeq
There, $\hat{a}_{1\mathrm{s}}^{\alpha/\beta}$ are the annihilation operators that remove one electron from the targeted core spin orbital and where the label ``$n$" designates the $n$\tsp{th} excited \dch state. Therefore, the intensity of a transition, $|T_{\textsc{i} \rightarrow \textsc{f},n}|^2$ (also known as pole strength or spectroscopic factor \cite{CedDomSch-PS-80,CedDomSch-ACP-86}), is proportional to the square of the overlap between the initial neutral wave function deprived of the two core electrons and the final \dch wave function. Note that the simplification of Eq.\,\eqref{overlap} can be connected to the sudden approximation as seen in Refs.~\cite{Abe-PR-67,MarShi-JCP-76}.

\subsection{The NOTA+CIPSI method}
\label{NOTA+CIPSI}

In this section we recall the characteristics of the NOTA+CIPSI method that we developed in ref.~\cite{FerPalPenIwaShiHikSoeItoLabTaiCar-JPCL-20}. This method was designed to provide answers suitable to the specificities of ssDCH spectroscopy for two key questions. First, how to compute the wave functions in equation~\eqref{overlap} ? Second, which set of molecular orbital (MO) to use for expanding these wave functions ?

Our answer to the first question is to use a selected configuration interaction method~\cite{BenDav-PR-69,HurMalRan-JCP-73,BuePey-TCA-74,BuePeyBru-BOOK-81,EvaDauMal-CP-83,Har-JCP-91} called Configuration Interaction using a Perturbative Selection made Iteratively (CIPSI). Such kind of method allows for an iterative building of the wave function's variational space by only adding important Slater determinants screened out via perturbation theory. Therefore, it provides a good description of the electronic relaxation and of the correlation effects and allows to compute enough excited states of the core dication while balancing their description. Thus, the expressions of the wave functions are 
\beq
\Ket{\Psi_\textsc{i}} = \sum_\mu C_\mu \Ket{\phi_\mu}
\eeq
where, $\Ket{\phi} = \Ket{1\mathrm{s}^\alpha,1\mathrm{s}^\beta,u_3,\cdots ,u_N}$ are $N$ electron Slater determinants in the \{$u$\} MO basis with a filled $1\mathrm{s}$ orbital in virtue of the frozen core approximation, and
\beq
\label{K-2_wf}
\Ket{\Psi_{\textsc{f},n}^{\tsc{k-2}}} = \sum_\nu C_\nu^n \Ket{\varphi^\textsc{\tsc{k-2}}_\nu}
\eeq
where $\Ket{\varphi^{\tsc{k-2}}} = \Ket{v_3,v_4,\cdots ,v_N}$ are $N-2$ electron Slater determinants in the \{$v$\} MO basis with an empty core orbital. This restriction on the double core hole location prevents the variational collapse of the method and arises from the core valence separation (CVS) approximation \cite{CedDomSch-PRA-80,Ced-PRA-87}. 

Answering the second question means defining the \{$u$\} and \{$v$\} MO sets. As removing two core electron from the system induces a strong relaxation of the electron cloud, we found that it is preferable to use two different sets respectively optimized for the initial neutral system and for the final ssDCH state. In practice, optimization of the MO basis set used in the computation of the neutral wave function, \{$u$\}, is performed via a restricted Hartree Fock (HF) calculation. Then, the relaxed \{$v$\} set is obtained by removing the two electrons from the targeted core MO and performing a biased SCF calculation that allows to partially re-optimize the \{$u$\} orbitals in the presence of the double core vacancy.  Different approaches exist to perform such reoptimization \cite{Bag-PR-65,JenJorAgr-JCP-87,AgrJen-CP-93,GilBesGil-JPCA-08}. In this work, we used an approach that keeps the double core hole orbital frozen in a similar fashion as suggested by Wood~\cite{Woo-CP-74}.

Using these two non mutually orthogonal MO sets is at the root of the NOTA+CIPSI method accuracy as we showed that it leads to a notably improved convergence of the spectrum~\cite{FerPalPenIwaShiHikSoeItoLabTaiCar-JPCL-20}. However, the computation of the overlap in \eqref{overlap} is non trivial as, according to L\"owdin's rule \cite{Low-PR-55}, the overlap between two Slater determinants is equal to the determinant of the overlap integrals between the two sets of occupied spin-orbitals.

\subsection{The CIPSI algorithm in a nutshell}
The CIPSI variant used in this work~\cite{QP-JCTC-19} uses a selection criterion based on second-order Epstein Nesbet perturbation theory \cite{Eps-PR-26,Nes-PRSA-55} computed via a semi stochastic method \cite{GarSceLooCaf-JCP-17}. The starting point of the procedure is a guess wave function $\Ket{\Psi_0}=\sum_{\mathrm{A}\in\mathcal{R}}C_\mathrm{A}\Ket{\mathrm{A}}$ where $\Ket{\mathrm{A}}$ represents Slater determinants included in the initial guess CI space $\mathcal{R}$. The importance of a determinant $\Ket{\mathrm{B}}$ non included in $\mathcal{R}$ is estimated through its contribution ${\cal E}_\text{B}^{(2)}$ to the second-order Epstein Nesbet perturbation theory energy,
\beq
{\cal E}_\text{B}^{(2)}= \frac{|\braket{\Psi_{0}|\hat{H}|\mathrm{B}}|^2}{ E_{0} -  \braket{\mathrm{B}|\hat{H} |\mathrm{B}}},
\eeq
where $\hat{H}$ is the electronic Hamiltonian and $E_0$ is the variational energy associated with the current iteration wave function. The determinants with the largest perturbative contribution are added to the CI space, $\mathcal{R}$ is updated, and the procedure is iterated until convergence is reached. In practice this selection procedure is made to balance the selection over all the states considered. At the end of the iterative procedure, the CIPSI energy is corrected for the remaining correlation energy using the renormalized second-order Epstein Nesbet perturbation theory scheme (rPT2),
\beq
\label{rpt2_correction}
\mathrm{E+}\mathrm{rPT2} = E_0 + Z \times \sum_\mathrm{B} {\cal E}_\text{B}^{(2)},
\eeq
with $Z\in [0\,;1]$ a renormalization factor which attenuates the magnitude of the correction for a small number of determinant (see ref.\,\onlinecite{QP-JCTC-19} for more details).

\section{Computational details}
\label{comput_details}
We used an home made plug-in for the software \tsc{quantum package~2.0}~\cite{QP-JCTC-19} that allows to compute NOTA+CIPSI spectra. 

The computational results presented on this paper focus on the O \dch spectra of the CO and \COO molecules. Details for CO were already given \cite{FerPalPenIwaShiHikSoeItoLabTaiCar-JPCL-20} and differ only from the following ones by the number of determinants included in the CI spaces. The NOTA+CIPSI results presented in this paper were obtained using the Dunning aug-cc-pVTZ (AVTZ) basis set \cite{Dun-JCP-89,KenDun-JCP-92}.

Regarding \COO, computation of the O \dch spectrum was performed with localized oxygen core MOs. Adopting a localized picture of the core hole is known to allow to treat a greater part of the relaxation following the core ionization at the MO level \cite{BagSch-JCP-72,CedDom-JCP-77,CedTarSch-JCP-87}. Therefore, such representation is coherent with the idea of the NOTA+CIPSI method which aims at using the most efficient set of MO for both initial and final states. Localizing the double core vacancy implies the breaking of the symmetry between the two possible double core ionization sites. Method like the non orthogonal CI approach \cite{BroNie-JMS-98,OosWhiGor-JCP-18} could prevent such symmetry breaking which however does not hinder the accuracy of our method.

The CIPSI procedure was carried out within the frozen core approximation for the initial neutral wave function and was iterated until the number of determinants was about $4.3 \times 10^6$. Furthermore, the double core hole was restricted to the core orbital of one of the two oxygen atoms during the CIPSI computation of the \dch wave functions, according to the CVS approximation. The initial guess used in the CIPSI procedure was the CI of single and double excitation (CISD) wave function which allows to start with an acceptable description of the ssDCH shake-up excited states. For both CO and \COO, the geometric relaxation of the final dication was neglected in our calculations, which is supported by our calculation on \COO including geometrical relaxation shown in section \ref{vib}

Because of the expression of the transition moment in Eq.\,\eqref{overlap}, monopolar selection rules apply, thus we only computed \dch states of the same symmetry as the neutral ground state. In the case of CO a total of 65 ssDCH states were considered for all presented spectra. But, for \COO, we had to compute around 200 states to reach the shake-up states responsible for the weaker satellite structure above 30~eV in the experimental spectrum (Fig.\,\ref{CO2_K-2_O}), which represents a heavy burden for any computational method. Moreover, this satellite region is dominated by transition to diffuse type shake-up states, as we showed previously in the case of CO (see \ref{CO_vs_CO2} and \cite{FerPalPenIwaShiHikSoeItoLabTaiCar-JPCL-20}), and is confirmed at the CISD level for \COO. Such states are of small interest for the following discussion as their smaller intensity makes it impossible to experimentally resolve the associated satellite structure. Therefore, we restricted our calculation on \COO to a total of 70 states allowing for a complete description of the three main structures displayed in the spectrum. We still show for reference the NOTA+CISD spectrum obtained with 250 ssDCH states in Fig.\,\ref{conv_CIPSI}.

To reduce the computation cost of equation \eqref{overlap} with two non mutually orthogonal basis, we applied a threshold to reduce the number of determinants to consider. In practice, this threshold was set such that we accounted for at least 99.9\% of the total norm of each states. Computation of the determinants in L\"owdin's expression of the transition matrix (see Refs.\,\onlinecite{FerPalPenIwaShiHikSoeItoLabTaiCar-JPCL-20} and~\onlinecite{Low-PR-55}) are performed using the LU decomposition approach. Finally all importants routines were parallellized using OpenMP. It may still be interesting to highlight that in the NOTA+CIPSI method, the main bottleneck remain by far the heavy wavefunction calculations.

The final \dch spectra were constructed using a Voigt profile. First, we used Lorentzian functions of area $|T_{\textsc{i} \rightarrow \textsc{f},n}|^2 $ and width $\Gamma_\tsc{f} = 0.5$~eV centered at the ionization energy associated with the corresponding final ssDCH state.  This value of the decay width corresponds to a rough estimate obtained as approximatively three times the decay width of the corresponding single core hole \cite{InhGroGru-JCP-13}. In the case of CO\tsb2, the lifetime width of the O single core hole was taken as 170~meV according to \cite{NicMir-JESRP-12}. Then, the whole spectrum was broadened by convolution with Gaussian functions of full width at half maximum equal to $3$~eV to account for the instrumental resolution that is limited by the need to detect the two photoelectrons in coincidence with a magnetic bottle spectrometer (see below).

Effects of geometric relaxation and vibronic transitions on the \COO spectrum profile were computed using the \tsc{gamess(us)} software at the Kohn-Sham DFT level of theory using the Becke \mbox{3-parameter} hybrid exchange \cite{Bec-JCP-93} and the Lee-Yang-Parr gradient-corrected correlation functional \cite{LeeYanPar-PRB-88} (B3LYP) and the \mbox{cc-pCVQZ} (CVQZ) basis set of Dunning \cite{Dun-JCP-89,WooDun-JCP-95}.

We also computed absolute O single site DCIP for CO and \COO using an adapted $\Delta \mathrm{CIPSI}$ approach (see \ref{new_section_DCIP} for more details). {Correlation effects arising from core electrons in the initial neutral system can not be neglected when aiming at computing absolute DCIP values therefore in this $\Delta \mathrm{CIPSI}$ core electron were correlated.} Identically, correction for relativistic effects is also non negligible. We estimated this correction at the CISD level using the third order Douglas--Kroll [DK(3)] transformation method applied to a $\Delta \mathrm{CISD}$ calculation using the software \tsc{gamess(us)}~\cite{GAMESS} and found a value of approximately 0.843~eV. Without surprise this value is fairly similar to the one that we previously computed for CO \cite{FerPalPenIwaShiHikSoeItoLabTaiCar-JPCL-20} reflecting the known atomic nature of such correction. We note that in the computation of the complete spectra, the differential effect of the relativistic correction between the ionization energy of different peaks of the same spectrum was neglected. 

Illustration of the electronic relaxation and shake-up induced reorganization in \COO are provided via two dimensional maps displaying the differences of densities integrated over the azimuth angle $\theta$ of a cylindrical basis $\{r, \theta, z\}$ where the z axis contains the three nuclei of the \COO molecule. The remaining $r$ and $z$ coordinates were discretized on a \mbox{$401\times401$} points grid. Similar representation of electronic reorganization induced by core ionization can be found in Refs.\,\cite{TasEhaCed-JCP-10,TakTasEhaUed-JPCA-11,TakYamUed-CPL-11,TakKryUed-JESRP-15,CorVazMar-JCTC-17,FerPalPenIwaShiHikSoeItoLabTaiCar-JPCL-20}. Molecular orbital profiles were obtained by using the interfacing between \tsc{quantum package~2.0} and the visualization software \tsc{Molden}~\cite{Molden1,Molden2}.

\section{Results and discussion}
\label{results}

\subsection{Accuracy tests of the NOTA+CIPSI method}
\label{convergence_section}

\subsubsection{Peak position convergence}

\begin{figure*}[t]
	\centering
	\includegraphics[scale=0.62]{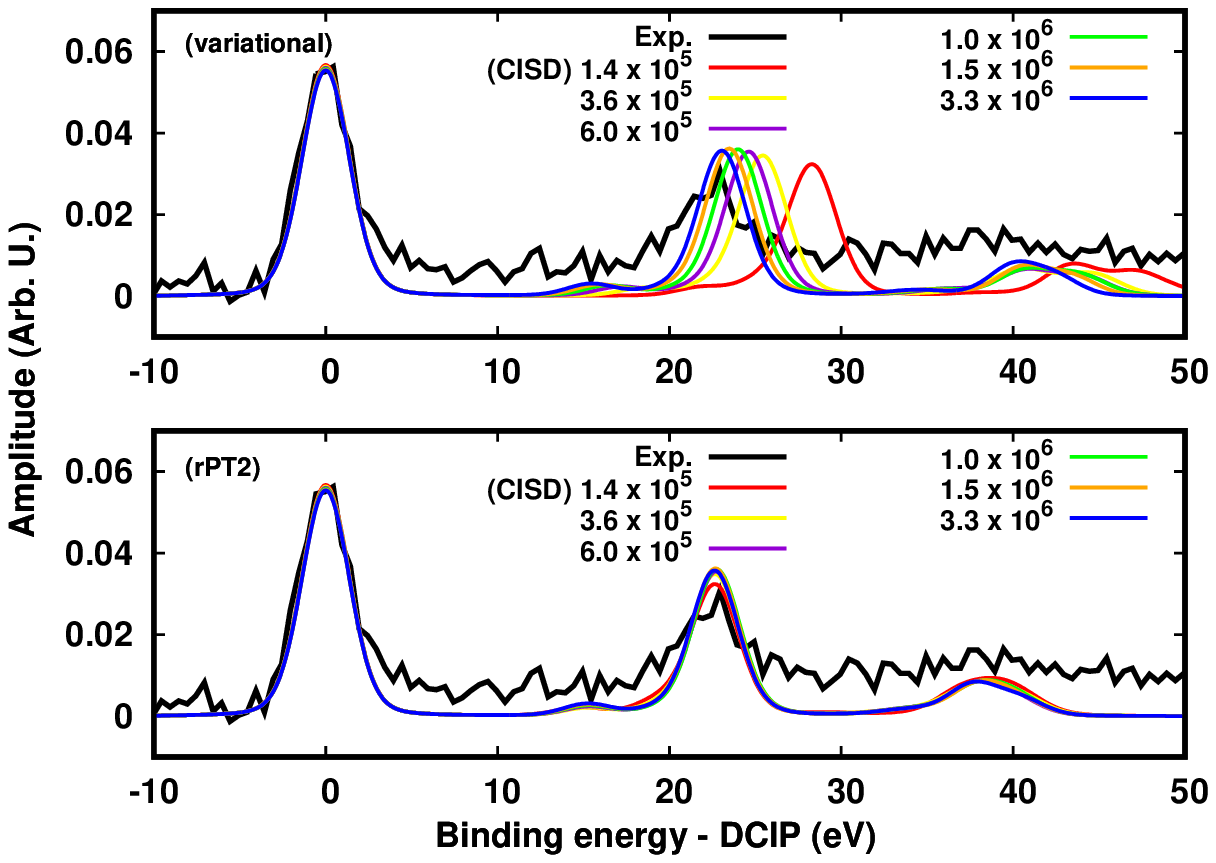}
	\includegraphics[scale=0.62]{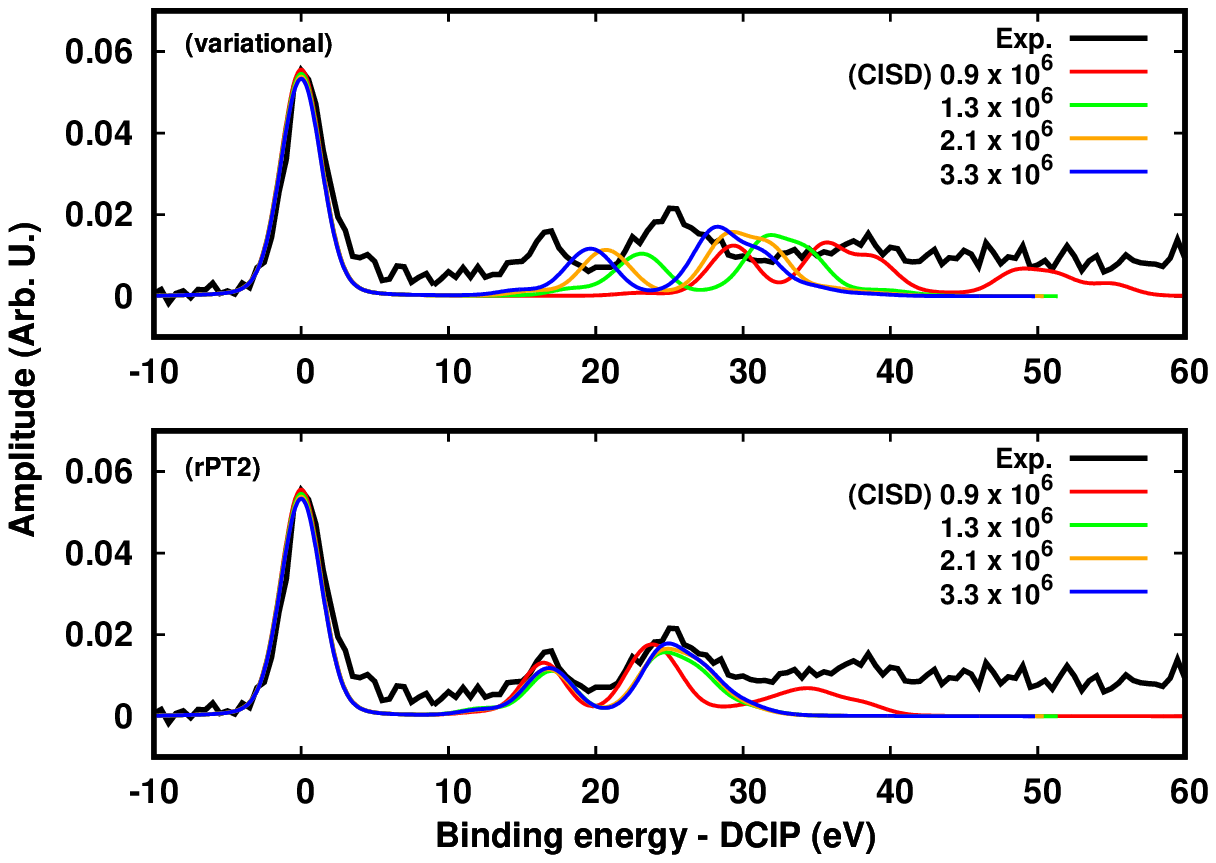}
	\caption{Convergence of the NOTA+CIPSI O \dch spectra of CO (left) and \COO (right) with respect to the number of determinants included in the \dch CI space. Top panels show the convergence when using the variational energy of the ssDCH wave functions while bottom ones depict the convergence of the spectra when applying the rPT2 correction to the core dication energies. For both CO and \COO, a unique well converged neutral wave function was used for all spectra. The CI expansions of these neutral wave functions contains respectively about 4.5 and 4.3 millions of selected Slater determinants.}
	\label{conv_CIPSI}
\end{figure*}

One of the parameters to take into account when choosing a computation method is its convergence behavior. Good convergence properties is what make the difference between reliable and uncertain results and so are of important computational interest. When using an iterative method, convergence with respect to the number of iterations, or, in the specific case of CIPSI, to the number of determinants, is especially important since it drastically affects the amount of time and of numerical resources that must be invested on the computation.

The convergence of NOTA+CIPSI spectra with respect to the number of selected Slater determinants included in the CI expansions of the final ssDCH states wave functions was investigated in the case of the O \dch spectra of the CO and \COO molecules. We report in figure \ref{conv_CIPSI}, the NOTA+CIPSI spectra of these two molecules obtained for a series of CIPSI iterations of the \dch wave function and using well converged neutral wave functions. Here, we focus our attention on the convergence of the relative position of the peaks, that is to say on the energy difference between the \dch ground state and each shake-up states. Therefore, we shifted all spectra to set the DCIP at 0~eV.

One of the issues encountered when looking at the convergence of such relative energies between main and shake-up satellite lines (which are intrinsically excitation energies) is that they are not variational quantities. Still, it is expected that for a small CI space, the position of the shake-up peaks with respect to the main line will be  overestimated if one only considers the variational energy of each states.  This is because in most of the cases, correlation effects are more important in the shake-up states than in the \dch ground state. This expected behavior can clearly be observed by focusing on the two non rPT2 corrected NOTA+CISD spectra (red curves) in the two top panels of Fig.~\ref{conv_CIPSI}. Indeed, for both CO and \COO the corresponding spectra display a clear shift of the satellite peaks toward the highest relative binding energies. 

As it is a non variational quantity, the evolution (with respect to the number of determinant) of the relative peak position is not intrinsically determined and we only know that it should reach the FCI limit at the end of the procedure. However, one can still observe a smooth and monotonous converging behavior of the NOTA+CIPSI spectra. Indeed, for both CO and \COO, as the CI space is enlarged, the shake-up peaks are gradually getting closer to the adequate position evidenced by the experimental spectra, and to the estimated FCI limit (see next paragraph).  This good convergence behavior is likely to be the result of a well balanced selection procedure over the different states considered, which may be greatly eased by their shared symmetry.   

Even though the computed spectra display such convergence, multiple iterations (thus a non negligible number of determinants) are still needed to accurately reproduce the experimental results. However, this is only true when looking at the non rPT2 corrected spectra.  If one now focuses on the two bottom panels of Fig.\,\ref{conv_CIPSI}, displaying the spectra obtained using the FCI estimates rPT2 corrected energies of eq.\,\eqref{rpt2_correction}, one sees that one or two CIPSI iterations were already enough to stabilize the peaks position for both molecules.

Here, we have left aside the question of convergence of transition amplitudes. This is because within the NOTA+CIPSI strategy, the use of two mutually non orthogonal molecular orbital basis sets yields to a very fast convergence for the intensities. Indeed, Fig.\,\ref{conv_CIPSI} only shows very little differences in the peak's amplitude when increasing the number of CIPSI iterations. The only noticeable difference is seen when comparing the initial CISD guess and the first CIPSI iterations. Such observation is not a surprise since, during the first iteration, the CI space is completed with the most important triply and quadruply excited Slater determinants. The primary effect of these determinants is to correct the CI coefficients associated with the connected singly and doubly excited determinants which are a determining factor in the intensity of the shake-up satellites (see \ref{CO2_satellites}, \ref{CO_vs_CO2} and Ref.\,\cite{FerPalPenIwaShiHikSoeItoLabTaiCar-JPCL-20}).

\begin{table*}[t]
\begin{threeparttable}

\renewcommand{\arraystretch}{1.8}
\setlength{\tabcolsep}{0.5cm}
\caption{CO and CO\tsb2 O single site DCIP and $\Delta$DCIP computed via the NOTA+CIPSI+DK(3) method and measured on different experimental set-ups. Bracketed values represent the difference between theory and experiment. All results are given in eV.}
\begin{tabular}{| c | c c c |}
	\hline	
     ~ & DCIP O K\tsp{$-2$} CO  & DCIP O K\tsp{$-2$} CO\tsb2\tnote{(a)} & $\Delta \mathrm{DCIP}$\tnote{(b)} \\ 
     \hline\hline
     \textbf{NOTA+CIPSI+DK(3)} & ~ & ~ & ~\\
     AVTZ / All electrons  & \textbf{1177.87 ($\mathbf{-0.13}$)} & \textbf{1174.06 ($\mathbf{-0.29}$)} & \textbf{3.81 $\mathbf{(+0.06)}$}\\
     \textbf{EXPERIMENTS} & ~ & ~ & ~\\
     This work (S.R.) & -  &  -  & 3.75 $\pm$ 0.4\hphantom{0}        \\  
     Ref.\,\onlinecite{LabCarIto-PRL-11} (S.R.) & 1178.0 $\pm$ 0.8 & 1174.7 $\pm$ 0.8 & 3.3 $\pm$ 1.6\\
     Ref.\,\onlinecite{SalUedPri-PRL-12} (X-FEL) & -  & 1173.2  $\pm$ 1.6  & - \\
     Extrapolated\tnote{(c)} & - & 1174.35 $\pm$ 0.45  & -\\ 
     \hline

\end{tabular}

\begin{tablenotes}
	\item[a] Computation errors are given with respect to the extrapolated \COO DCIP.
	\item[b] Computation errors are given with respect to the $\Delta \mathrm{DCIP}$ measured in this work.
	\item[c] Represents the common confidence interval between the two experimental results. 
\end{tablenotes}

\label{table_DCIP}

\end{threeparttable}
\end{table*}

\subsubsection{Double core ionization potential}
\label{new_section_DCIP}

Up to this point, we only discussed relative binding energies as we focused our attention on the position and intensity of the peaks in a given spectrum. However, the absolute value of the DCIP of a system as well as the difference of binding energy between two molecules are pieces of information that hold as much spectroscopic interest.

Experimental measurement of DCIP are usually associated with an important uncertainty ranging from about half an eV to few eVs depending on the system and set-up. This is mainly due to the combination of the experimental resolution with the calibration error. However, this last source of uncertainty can be nullified by considering relative DCIP. Indeed, if the measurements are performed in identical experimental conditions, the calibration error will cancel by itself and conducts to an enhanced precision. While it is trivial when comparing signals within the same spectrum, it is more challenging when looking at different systems. 

In the present case, the O ssDCH spectra of CO and CO\tsb2 were recorded within the same experimental conditions. For the whole experiment duration, the photon energy was kept fixed at 1341~eV while different target gases were studied successively. The same gas was also studied at different time to check a possible drift of the monochromator. Due to the electron energy resolution ($\sim$\,3~eV) the precision in the relative energy shift between different target gases can be given with 400 meV confidence taking into account possible change in contact potentials when using different target gases  which are however not very reactive. Therefore, we have a reliable measure of the difference between the two O DCIP of CO and CO\tsb2 (henceforth dubbed as $\Delta\mathrm{DCIP}$) which we determined to be 3.75 $\pm$ 0.4 eV.

We computed the O \dch DCIP for the CO and \COO molecules using the NOTA+CIPSI+DK(3) strategy, \textit{i.e.} we performed {an independent} $\Delta\mathrm{CIPSI}$ calculation using the two MO basis sets defined in \ref{NOTA+CIPSI} and corrected by the DK(3) relativistic correction (see~\ref{comput_details}). 
For both molecules, the CIPSI procedures were iterated until the number of determinants was superior to $12 \times 10^6$. All electrons were correlated including the core ones in the neutral systems. Our results, are compared to our new experimental data and to data from the literature which are all compiled in table~\ref{table_DCIP}. In the case of the \COO DCIP, two experimental results were available, one obtained using synchrotron radiation (S.R.), the other via \mbox{X-FEL}. Thus, in order to be as objective as possible we compare our result to the median value of the shared confidence interval between the two experiments.

The predicted values obtained with our NOTA+CIPSI+DK(3) method are in good agreement for both the DCIP of CO and \COO as well as for their relative $\Delta$DCIP. Regarding these two molecules, other methods had already been applied to compute their binding energies \cite{TasEhaCed-JCP-10,LeeSmaGor-JCP-19,LabCarIto-PRL-11,TasUedEha-CPL-11,UedTak-JESRP-12} and while some of them predicted satisfactory values for either the absolute DCIP or for the relative one, none  was able to satisfactory yield results within the experimental error bars for both quantities. The accurate predictions of the NOTA+CIPSI+DK(3) method for both DCIP and $\Delta$DCIP values result from its good description of relaxation and correlation within the neutral and the \dch ion, as well as its good control of the error between different molecular systems.

\subsection{Detailed interpretation of the O single site double core hole spectrum of CO\tsb{2}}
\label{CO2}

\subsubsection{General shape of the spectrum}

\begin{figure}[b]
        \centering
        \includegraphics[scale=0.62]{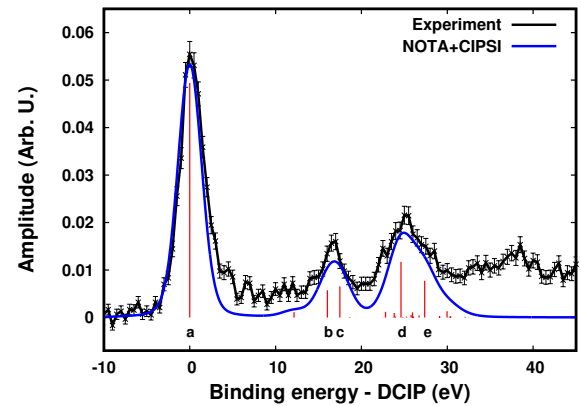} 
        \caption{Experimental (black) and computed NOTA+CIPSI (blue) O \dch spectrum of CO\tsb{2}. Red lines are of size $0.25\times |T_{\textsc{i} \rightarrow \textsc{f},n}|^2$. 
        Only state with a transition amplitude superior to 0.01 were given a \mbox{label.}}
        \label{CO2_K-2_O}
\end{figure}

We report in Fig.\,\ref{CO2_K-2_O} the O \dch spectrum of the CO\tsb{2} molecule computed via the NOTA+CIPSI method alongside its experimental counterpart. In the aforementioned figure, each spectrum was individually shifted such that the maximum of the main peak is set to 0~eV. The intensity of the most intense transitions, the binding energy of the associated ssDCH states and the main components of their CI expansions are compiled in table~\ref{table_CO2}. We also show in Fig.\,\ref{mo_CO2} the profile of the relaxed \dch optimized \{$v$\} MOs strongly involved in the descriptions of these states. To distinguish between the two oxygen centers in the \COO O \dch dication, we use the $\dag$ symbol to mark the atom bearing the double core vacancy. As an example, the subscript ``CO$^\dag$" indicates that the associated orbital is mainly localized on the carbon atom and on the hollow oxygen atom.

 The presented experimental O \dch spectrum of \COO was measured using a magnetic bottle time-of-flight (MB-TOF) electron spectrometer on the BL-16 undulator beamline at the Photon Factory and have improved statistics in comparison to Ref.\,\cite{LabCarIto-PRL-11}. It was recorded within identical experimental conditions as the CO spectrum that was measured at the same time (see Fig.\,\ref{conv_CIPSI} and ref.\,\cite{FerPalPenIwaShiHikSoeItoLabTaiCar-JPCL-20}) and the experimental data were analyzed using the same \mbox{three-electron} coincidence approach. For the photon energy used, 1341 eV, the total kinetic energy carried by the two photo electrons is hence about 170~eV and gives an experimental energy resolution of about 3~eV. These corresponds to a relative resolution $\Delta\mathrm{E}/\mathrm{E}$ of about 2\%, that we have with the magnetic bottle.

The NOTA+CIPSI method very accurately reproduces the three main features of the CO\tsb{2} O \dch spectrum both in terms of relative intensity and position of the peaks. The first and most intense peak of the spectrum stems from the transition to the fully relaxed \dch ground state. Both the peak and the related final state are labeled ``\textit{a}" in figure~\ref{CO2_K-2_O} and table~\ref{table_CO2}. Two pronounced satellite regions around 14--20~eV and 21--31~eV with respect to the DCIP are both shown to be mostly the consequence of the transition to two shake-up states, \textit{b} and \textit{c} for the first one and~\textit{d} and~\textit{e} for the second. Then, follows a third broad and less marked satellite region around 32--42 eV whose origin was traced to diffuse type shake-up states via a NOTA+CISD calculation accounting for 250 \dch states (see Fig.\,\ref{conv_CIPSI} for the corresponding spectrum).

\subsubsection{Influence of vibrational effects and geometric relaxation}
\label{vib}
 Following the removal of two core electrons from the same atomic center, geometric relaxation of the molecule is expected {\cite{DomCed-PRA-77,CedDomSch-PRA-80}}. We present thereafter our computational investigation of these effects and show that, in the present case, they remain negligible in front of the experimental resolution.
 
 Geometry optimization and harmonic frequencies calculation (KS-DFT B3LYP/CVQZ see section~\ref{comput_details}) were performed for both neutral and O ssDCH  electronic ground states of \COO. Table~\ref{GS_geom} summarizes these results. We note that in both Tables~\ref{GS_geom} and \ref{table_fc} we only indicated values for one of the two bending modes. As one can see, the molecular geometry is substantially affected by the double core vacancy. In particular, the CO$^\dag$ bond length is increased by about 0.2 \AA{} while the second OC bond is reduced by 0.06 \AA. Consequently one can expect a relative shift in binding energy compared to the vertical energy differences at the neutral ground state geometry. Moreover, since the \dch potential energy surface is likely steep at the neutral equilibrium geometry a broadening of the spectrum should be observed.
 
\begin{table}[h]

\centering
\renewcommand{\arraystretch}{1.4}
\setlength{\tabcolsep}{0.5cm}
\caption{KS-DFT optimized geometries of the neutral and O ssDCH of \COO. Distances are given in {\AA} and frequencies in cm$^{-1}$.}
\begin{tabular}{ccc}\hline\hline
  &   \multicolumn{1}{c}{Neutral} &  O \dch \\\hline
    d$_{\tsc{CO}^\dag}$     & 1.160 & 1.369\\
    d$_{\tsc{CO}}$          & 1.160 & 1.104\\
     $\nu_{\mathrm{bend}}$\tnote{(a)}  & 660   & 376 \\
     $\nu_{\mathrm{sym}}$   & 1364  & 848 \\
     $\nu_{\mathrm{asym}}$  & 2386  & 2444 \\\hline\hline
\end{tabular}

\label{GS_geom}
\end{table}

The vertical (DK(3) corrected) DCIP value is estimated at 1171.21 eV within our (KS-DFT B3LYP/CVQZ) calculation. As observed this value is underestimated compared to the experiment and the NOTA+CIPSI+DK(3) ones. This is not surprising since DCH calculations at DFT level of theory suffer of a systematic underestimation of about 2 to 3 eV compared to exact values \cite{CarSelFerBerWuoNakHikItoZitBucAndPalPenLab-JCP-19}. On the other hand, the adiabatic binding energy is found equal to 1169.90 eV indicating a negative shift of about $-1.3$ eV due to geometric relaxation in the ssDCH system.

Vibrational profile of the main peak was computed within the Franck-Condon approximation while neglecting temperature effects. We assumed that the transition dipole moment associated with vibronic transitions can be reduced to the overlap between initial and final vibrational wave functions,
\beq
\label{overlap_vib}
T_{\mathbf{\boldsymbol{\nu}} \rightarrow \boldsymbol{\nu^\prime}} \propto \Braket{\Psi_{\boldsymbol{\nu^\prime}}^{\tsc{f}} |  \Psi_{\boldsymbol{\nu}}^{0}},
\eeq
where  $\Ket{\Psi_{\boldsymbol{\nu}}^{0}}$ and   $\Ket{\Psi_{\boldsymbol{\nu^\prime}}^{\tsc{f}}}$ are the total vibrational wave functions associated to the neutral and to the final ssDCH electronic ground states respectively and $\boldsymbol{\nu} = (v_{\mathrm{bend,1}}, v_{\mathrm{bend},2}, v_\mathrm{sym}, v_\mathrm{asym})$ and ${\boldsymbol{\nu^\prime} = (v_{\mathrm{bend},1}^\prime, v_{\mathrm{bend},2}^\prime, v_\mathrm{sym}^\prime, v_\mathrm{asym}^\prime)}$ are vectors of vibrational quantum numbers associated to the \mbox{($3N_\mathrm{nuc}-5=4$)} normal modes in the initial and final electronic states respectively.

Overlap in eq.\,\eqref{overlap_vib} were computed assuming that initial and final total vibrational wave functions are developed as a product of vibrational wave functions corresponding to each of the initial normal modes,
\beq
\Psi_{\boldsymbol{\nu}}^{0} (\textbf{R}) = \prod_{k=1}^{3N_\mathrm{nuc}-5} \mathcal{V}_{v_k}(Q_k-Q_{0,k}^\mathrm{eq};\omega_k),
\eeq
and
\beq
\label{vib_no_rot}
\Psi_{\boldsymbol{\nu^\prime}}^{\tsc{f}}(\textbf{R}^{\boldsymbol{\prime}})  = \prod_{l=1}^{3N_\mathrm{nuc}-5} \mathcal{V}_{v^\prime_l}(Q^\prime_l-Q_{f,l}^{\prime\,\mathrm{eq}};\omega^\prime_l),
\eeq
where,  $\mathcal{V}_v(Q-Q^\mathrm{eq};\omega)$ represents a normalized eigenfunction of a harmonic oscillator of reduced mass $\mu$ corresponding to the associated normal mode, centered at the normal coordinate $Q^\mathrm{eq}$, in a vibrational quantum number $v$ and undergoing vibration along the $Q$ coordinate with frequency~$\omega$.

Normal modes and vibrational frequencies of the \dch dication should differ from those of  the neutral system, a phenomenon first considered by Duschinsky \cite{Dus-USSR-37} when extending
the Franck-Condon principle from diatomics to polyatomic molecules. We considered the displacement of the equilibrium geometry between neutral and \dch state \mbox{($Q_{0,k}^\mathrm{eq} \neq Q_{f,k}^{\prime\,\mathrm{eq}} $)}. However, we fully neglected Duschinsky's rotation of the normal modes in eq.\,\eqref{vib_no_rot} as we found that the projection of the initial normal modes onto their final counterpart are all fairly close to $1$ (0.99, 0.94 and 0.94 for bending, symmetric stretching and asymmetric stretching, respectively).

Franck-Condon factors have been computed using the expression of the overlap integrals derived by Katriel \cite{Kat-JPB-70},
\begin{align}
\label{katri}
&I(\omega'_{k},v'_{k};\omega_{k},v_{k})=\nonumber\\
&\left[(v_{k}!\,v'_{k}!)  \left({\frac{2(\omega'_{k}\,\omega_{k})^\frac{1}{2}}{\omega{'}_{k}+\omega_{k}}}\right) \right]^\frac{1}{2}\,\exp\left({-S^{2}_{k}\frac{\omega{'}_{k}}{\omega{'}_{k}+\omega_{k}}}\right) \nonumber \times \\
&\sum_{(\mathrm{i}_{1},\mathrm{i}_{2},\mathrm{i}_{3})}\Bigg(\frac{1}{\mathrm{i}_{1}!\,\mathrm{i}_{2}!\,\mathrm{i}_{3}!\,(v_{k}-2\mathrm{i}_{1}-\mathrm{i}_{3})!\,(v'_{k}-2\mathrm{i}_{2}-\mathrm{i}_{3})!} \nonumber\\
&\times\left[\frac{\omega_{k}-\omega'_{k}}{2(\omega'_{k}-\omega_{k})}\right]^{\mathrm{i}_{1}}
\left[\frac{\omega'_{k}-\omega_{k}}{2(\omega'_{k}+\omega_{k})}\right]^{\mathrm{i}_{2}}
\left[{\frac{2(\omega'_{k}\,\omega_{k})^{1/2}}{\omega'_{k}+\omega_{k}}}\right]^{\mathrm{i}_{3}}\nonumber\\
&\times\left[2S_{k}\left(\frac{\omega'_{k}}{\omega_{k}}\right)^\frac{1}{2} \frac{\omega_{k}}{(\omega'_{k}+\omega_{k})}\right]^{(v'_{k}-2\,\mathrm{i}_{2}-\mathrm{i}_{3})}\nonumber\\
&\times\left[-2 S_{k} \frac{\omega'_{k}}{(\omega'_{k}+\omega_{k})}\right]^{(v_{k}-2\,\mathrm{i}_{1}-\mathrm{i}_{3})}\Bigg),
\end{align}
where $(\mathrm{i}_{1,2,3})$, $(v_{k}-2\,\mathrm{i}_{1}-\mathrm{i}_{3})$ and $(v'_{k}-2\,\mathrm{i}_{2}-\mathrm{i}_{3})$ are all non-negative integers. This procedure permits to consider Franck-Condon profile for normal mode in which the potential energy derivative, with respect to a given coordinate, is zero by symmetry.

In the above equation \eqref{katri}, we introduced a unitless vibronic coupling constant, denoted here ($S_k$), which is equivalent to the $\Big(\sqrt{b\frac{M\Omega}{2\hbar}}\,\Big)$ term in Katriel's paper. We followed the steps of Ref.\,\onlinecite{DarSorSve-JCP-98} to estimate ($S_k$) which is the measure of difference in equilibrium geometries between the ground state and the final core excited state. From this procedure it comes that,
\beq
S_{k}=\Delta{Q}_{k}\sqrt{\frac{\omega_{k}}{2{\hbar}}},
\eeq
where, $\Delta Q_{k}$ represents the change in normal coordinate~$k$ in M$^{(1/2)}\,$L unit. This can be computed using \tsc{gamess(us)} output as,
\beq
\Delta{Q}_{k}=\sum_{j=1}^{3N_{\mathrm{nuc}}} x_{k,j} m_{j}(X_{f}(j)-X_{o}(j)),
\label{tom}
\eeq
where $X_{f}(j)-X_{o}(j)$ is the difference between the $j$\tsp{th} Cartesian coordinates of the ionic state and the neutral ground state, 
$x_{k,j}$ is the cartesian coordinate of the normal mode and $m_j$ is the mass of the atom associated with the $j$\tsp{th} coordinate.

\begin{figure}[t]
		\centering 
		\includegraphics[scale=0.6]{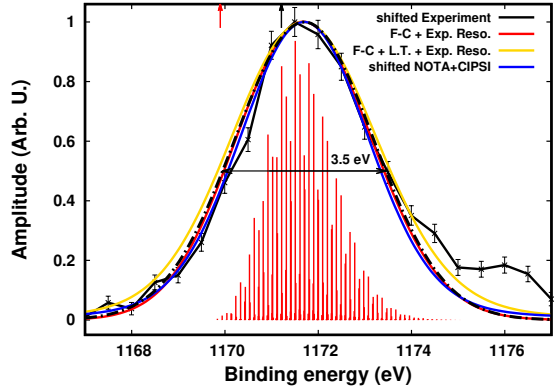}
        \caption{Computed vibrational profile of the main peak of the O ssDCH spectrum of \COO. Red lines represent the Franck-Condon intensities. Red curve is the associated profile taking into account the experimental resolution of 3~eV. Gold curve also takes into account the lifetime broadening of 0.5~eV. Blue curve is the (shifted) NOTA+CIPSI vertical spectrum. Plain black curve is the (shifted) experimental profile. Dashed black curve is a Gaussian function of FWHM=3.5 eV centered at 1171.70~eV. Red and black arrows indicates respectively the DK(3) corrected KS-DFT B3LYP/CVQZ  adiabatic and vertical DCIP.}
        \label{fig_vib}
\end{figure}

Franck-Condon factors can then be computed as the square of the above overlap integrals \eqref{katri}, for any ($\omega{'}_{k},v'_{k};\omega_{k},v_{k}$) terms. We report in Table~\ref{table_fc} the Franck-Condon values for every $(\omega^\prime_k,v^\prime_k,\omega_0,v_0)$ terms and show, in Figure~\ref{fig_vib}, the corresponding theoretical vibrational profile associated to the main peak of the \COO O \dch spectrum which is compared to the profile obtained at the vertical NOTA+CIPSI level (we also provide as Supplementary Materials the vibrational profiles of the main peak of the O and C K$^{-1}$ spectrum of CO as a proof of concept for this approach of the computation of vibronic contributions).

\begin{table*}[t]
\centering
\renewcommand{\arraystretch}{1.7}
\caption{Vertically: KS-DFT B3LYP/CVQZ frequencies of the harmonic vibrational modes of the ssDCH final state; Horizontally: Franck-Condon values for up to the 16th vibrational levels.}
\resizebox{\textwidth}{!}{
\begin{tabular}{|c||c||c|c|c|c|c|c|c|c|c|c|c|c|c|c|c|c|c|}\hline
  M & Freq. (cm$^{-1}$)   &      $v_{0}$                 &     $v_{1}$     &      $v_{2}$  &  $v_{3}$      &   $v_{4}$     &    $v_{5}$   &   $v_{6}$    &   $v_{7}$     &    $v_{8}$   &  $v_{9}$       &   $v_{10}$      &   $v_{11}$             &   $v_{12}$       &      $v_{13}$   &          $v_{14}$ &     $v_{15}$     &     $v_{16}$      \\\hline  \hline 
  $\nu_{\mathrm{bend}}$\tnote{(a)}  &     376     &  0.961&  0.000&  0.036  &0.000  &0.002 & 0.000&  0.0001& &&&&&&&&&\\
  $\nu_{\mathrm{sym}}$   &    848    &0.0538&  0.1921&  0.2995&  0.2637& 0.1402&  0.0438&  0.0067&  0.0002&  0.0001 & 0.0001 & 0.0000&&&&&& \\
  $\nu_{\mathrm{asym}}$  &     2444   &0.0042&  0.0228&  0.0619&  0.1124&  0.1538& 0.1691&  0.1556&  0.1233&  0.0858&  0.0533&  0.0299&  0.0154&  0.0072&  0.0032&  0.0013&  0.0005&  0.0002  \\\hline
\end{tabular}
}
\label{table_fc}
\end{table*}

Taking into account the vibrational profile and the experimental resolution of 3 eV, the maximum of the peak is found at about 1171.70 eV. The 1.8 eV positive shift with respect to the adiabatic DCIP value is mainly due to the activation of the asymmetric stretching mode of \COO and more moderately from the symmetric stretching mode. This is not surprising since the relaxed geometry of the \dch dication exhibits two very different bond lengths.  As a result, the maximum of the Franck-Condon associated to the asymmetric mode is found for $v=5$ while only few modes are activated for the symmetric one (maximum for $v=2$).

Because of compensation between the negative and positive energy shifts respectively due to geometrical relaxation and vibronic transitions, the maximum of the peak is found very close to the \mbox{KS-DFT} vertical DCIP value. Also, the vibrational profile is only slightly assymetric therefore the Franck-Condon {full width at half maximum (FWHM)} is estimated to $\approx 1.5$ eV and yields a total {FWHM} of  $\approx$~3.5~eV. Despite vibrational effects and geometrical relaxation, the final profile taking into account life-time broadening and experimental energy resolution (gold curve) does not change significantly the final width of the peaks when compared to the profile obtained by only considering vertical transitions and neglecting vibronic contributions (blue curve). 

As we only considered the main peak these observations can not be straightforwardly extended to shake-up peaks, still, we strongly expect that the conclusion should remain the same.  Therefore, and as the following discussion will focus on electronic properties we will discard these effects for the rest of the paper.

\subsubsection{\dch ground state and double core hole induced relaxation}
\label{relaxation_K-2}

The monopolar form of the transition moment in eq.\,\eqref{overlap} indicates that the more similar the final dication wave function is to the initial neutral wave function, the higher the associated transition probability. The main difference between initial and final system is the strong electronic relaxation caused by the formation of the double core vacancy. Nevertheless, the \dch ground state still retains some similarity with the neutral ground state as evidenced by the relatively high intensity of the main peak of the spectrum.


\begin{table*}[p]
\begin{threeparttable}
\renewcommand{\arraystretch}{1.9}
\setlength{\tabcolsep}{0.25cm}
\caption{NOTA+CIPSI transition amplitudes and relative binding energies associated with the intense peaks of the O \dch spectrum of CO\tsb2, and their corresponding final ssDCH state.}
\begin{tabular}{| c c c p{10.5 cm} |}
	\hline	
    Label\tnote{(a)} & (BE$-$DCIP) & $|T_{\textsc{i} \rightarrow \textsc{f},\mathrm{n}}|^2$ & Main configurations\tnote{(b,c,e)} \\
    \hline \hline
    \textit{a} & 0        & 0.198 &  $-\,0.93$\ \ \ \, $\big\{(1\,\mathrm{s}_{\tsc{o\tsp{\dag}}})^0\,\big\}$ \\
    \textit{b} &  16.030  & 0.023 &  
    $-\,0.33$ $\big\{(1\,\mathrm{s}_{\tsc{o\tsp{\dag}}})^0\,(2'\pi_{\tsc{oc}})^0\,(3'{\pi^*})^2\,\big\}\,$
    $-\,0.33$ $\big\{(1\,\mathrm{s}_{\tsc{o\tsp{\dag}}})^0\,(2\,\pi_{\tsc{oc}})^0\,(3\,{\pi^*})^2\,\big\}$
    $+\,0.26$ $\big\{(1\,\mathrm{s}_{\tsc{o\tsp{\dag}}})^0\,(2\,\pi_{\tsc{oc}}^{\alpha})^1\,(2'\pi_{\tsc{oc}}^{\alpha})^1\,(3\,{\pi^*}^{\beta})^1\,(3'{\pi^*}^{\beta})^1\,\big\}$ \hphantom{************.***}
    $-\,0.26$~\,\,\,\,~$\big\{(1\,\mathrm{s}_{\tsc{o\tsp{\dag}}})^0\,(2\,\pi_{\tsc{oc}}^{\alpha})^1\,(2'\pi_{\tsc{oc}}^{\beta})^1\,(3\,{\pi^*}^{\alpha})^1\,(3'{\pi^*}^{\beta})^1\,\big\}$ \hphantom{*************.****}
    $-\,0.24$\,\,\,\,\,\,\,\,$\big\{(1\,\mathrm{s}_{\tsc{o\tsp{\dag}}})^0\,(4\,\sigma_{\tsc{oc}}^*)^1\,(5\,{\sigma})^1\,\,\,\,\,\,\big\}$\, 
    $+\,0.20$~$\big\{(1\,\mathrm{s}_{\tsc{o\tsp{\dag}}})^0\,(2\,\pi_{\tsc{oc}})^1\,(3\,{\pi^*})^1\,\big\}$
    $+\,0.20$\,\,\,\,\,\,\,\,$\big\{(1\,\mathrm{s}_{\tsc{o\tsp{\dag}}})^0\,(2'\pi_{\tsc{oc}})^1\,(3'{\pi^*})^1\,\,\big\}$\,\,\,\,\,\,\,
    $-\,0.11$~\hspace{+0.33 cm}$\big\{(1\,\mathrm{s}_{\tsc{o\tsp{\dag}}})^0\,\big\}$
     \\
    \textit{c} &  17.486  & 0.026 &  
    $+\,0.30$ $\big\{(1\,\mathrm{s}_{\tsc{o\tsp{\dag}}})^0\,(2'\pi_{\tsc{oc}})^1\,(3'{\pi^*})^1\,\big\}$
	$+\,0.30$ $\big\{(1\,\mathrm{s}_{\tsc{o\tsp{\dag}}})^0\,(2\,\pi_{\tsc{oc}})^1\,(3\,{\pi^*})^1\,\big\}$
	$-\,0.23$ \hspace{+0.1 cm}$\big\{(1\,\mathrm{s}_{\tsc{o\tsp{\dag}}})^0\,(2\,\pi_{\tsc{oc}}^{\alpha})^1\,(2'\pi_{\tsc{oc}}^{\alpha})^1\,(3\,{\pi^*}^{\beta})^1\,(3'{\pi^*}^{\beta})^1\,\big\}$ \hphantom{****************}
	$+\,0.23$\tnote{(d)}~\,\,\,~$\big\{(1\,\mathrm{s}_{\tsc{o\tsp{\dag}}})^0\,(2'\pi_{\tsc{oc}})^0\,(3'{\pi^*})^2\,\big\}$\, \,\,
	$+\,0.22$\tnote{(d)}~\,\,\,~$\big\{(1\,\mathrm{s}_{\tsc{o\tsp{\dag}}})^0\,(2\,\pi_{\tsc{oc}})^0\,(3\,{\pi^*})^2\,\big\}$
	$-\,0.21$ \, \,\,$\big\{(1\,\mathrm{s}_{\tsc{o\tsp{\dag}}})^0\,(4\,\sigma_{\tsc{oc}}^*)^1\,(5\,{\sigma})^1\,\,\,\,\big\}$ \hphantom{******************.....************}
	$+\,0.17$ \hspace{+0.15 cm}$\big\{(1\,\mathrm{s}_{\tsc{o\tsp{\dag}}})^0\,(2\,\pi_{\tsc{oc}}^{\alpha})^1\,(2'\pi_{\tsc{oc}}^{\beta})^1\,(3\,{\pi^*}^{\alpha})^1\,(3'{\pi^*}^{\beta})^1\,\big\}$ \hphantom{****************}
	$\,\,-\,0.17$\tnote{(d)} \ \ \, $\big\{(1\,\mathrm{s}_{\tsc{o\tsp{\dag}}})^0\,(2\,\pi_{\tsc{oc}})^0\,(3'{\pi^*})^2\,\,\big\}$\,\, \,\,\,
	$-\,0.16$\tnote{(d)} \,\,\,\, $\big\{(1\,\mathrm{s}_{\tsc{o\tsp{\dag}}})^0\,(2'\pi_{\tsc{oc}})^0\,(3\,{\pi^*})^2\,\big\}$
     \\
    \textit{d} &  24.603  & 0.047 &  
    $-\,0.38$ \hspace{+0.37 cm}$\big\{(1\,\mathrm{s}_{\tsc{o\tsp{\dag}}})^0\,(1\,\pi_{\tsc{co\tsp{\dag}}})^1\,(3'{\pi^*})^1\,\big\}$\hspace{+0.30 cm}
    $+\,0.38$ \hspace{+0.33  cm}$\big\{(1\,\mathrm{s}_{\tsc{o\tsp{\dag}}})^0\,(1\,\pi_{\tsc{co\tsp{\dag}}})^1\,(3\,{\pi^*})^1\big\}$
    
    $-\,0.25$ \hspace{+0.34 cm}$\big\{(1\,\mathrm{s}_{\tsc{o\tsp{\dag}}})^0\,(3\,\sigma_{\tsc{co\tsp{\dag}}}^*)^1\,(5\,{\sigma})^1\,\ \, \big\}$
    \\
    \textit{e} &  27.367  & 0.031 &  $+\,0.55$ \hspace{+0.35 cm}$\big\{(1\,\mathrm{s}_{\tsc{o\tsp{\dag}}})^0\,(3\,\sigma_{\tsc{co\tsp{\dag}}}^*)^1\,(5\,{\sigma})^1\hspace{+0.22 cm}\big\}$ \\
    \hline
\end{tabular}

\begin{tablenotes}
	\item[a] Only states with $|T_{\textsc{i} \rightarrow \textsc{f},\mathrm{n}}|^2 > 0.01$
	\item[b] Only configurations with CI coefficients in intermediate normalization $C_\nu^n / \mathrm{max}\{C^n_\nu\} > 0.3 $
	\item[c] Open shell configurations are only reported once but appear twice in the expansion because of the $\alpha$-$\beta$ symmetry.
	\item[d] This spurious symetry breaking is here strongly overamplified by means of the rounding. We believe it to originate from a too lose value of the threshold used to enforce the symmetry which leads to the inappropriate discarding of some determinants when the CI space became too large.
	\item[e] Show occupancy differences with respect to the reference configuration : \\
	$(1\,\mathrm{s}_{\tsc{o\tsp{\dag}}})^2$\,$(1\,\mathrm{s}_{\tsc{o}})^2$\,$(1\,\mathrm{s}_{\tsc{c}})^2$\,$(1\,\sigma_{\tsc{co\tsp{\dag}}})^2$\,$(2\,\sigma_{\tsc{oc}})^2$\,$(3\,\sigma_{\tsc{co\tsp{\dag}}}^*)^2$\,$(1\,\pi_{\tsc{co\tsp{\dag}}})^2$\,$(1'\pi_{\tsc{co\tsp{\dag}}})^2$\,$(4\,\sigma_{\tsc{oc}}^*)^2$\,$(2\,\pi_{\tsc{oc}})^2$\,$(2'\pi_{\tsc{oc}})^2$
\end{tablenotes}

\label{table_CO2}

\end{threeparttable}
\end{table*}
\begin{figure*}[p]
		\centering 
		\includegraphics[scale=0.36]{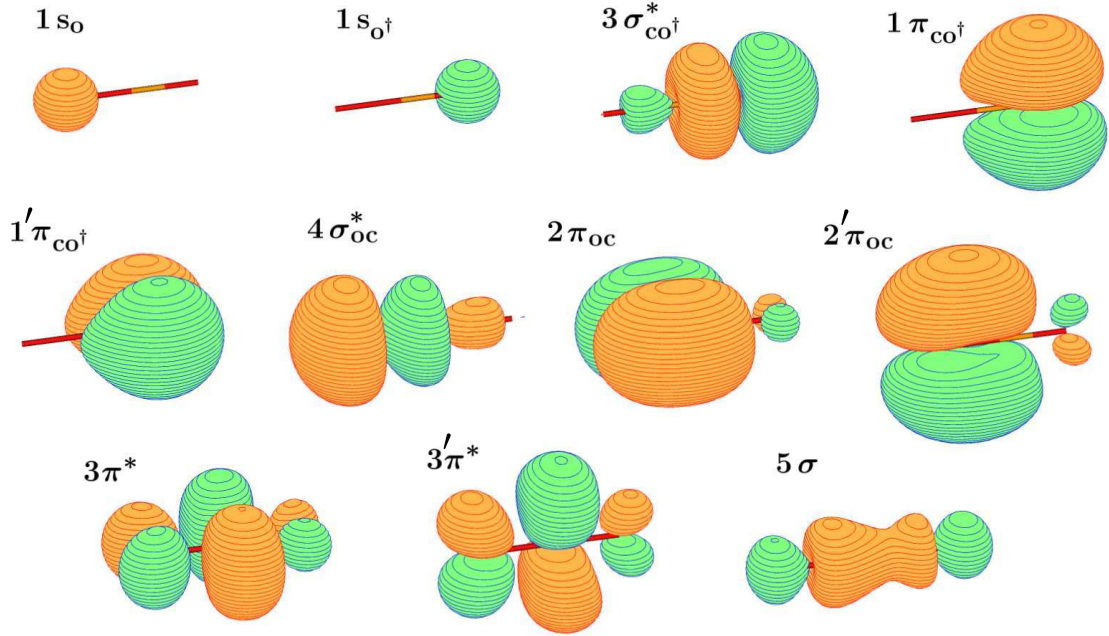}
        \caption{Profile of the relaxed, \dch optimized, \{$v$\} molecular orbitals strongly involved in the descriptions of the shake-up states associated with intense transition in the O ssDCH spectrum of \COO. The MOs have been plotted for a contour value of $0.2$. First two rows show the occupied MOs while the last one shows the virtual ones.}
        \label{mo_CO2}
\end{figure*}

\begin{figure}[b]
		\centering 
		\includegraphics[scale=0.38]{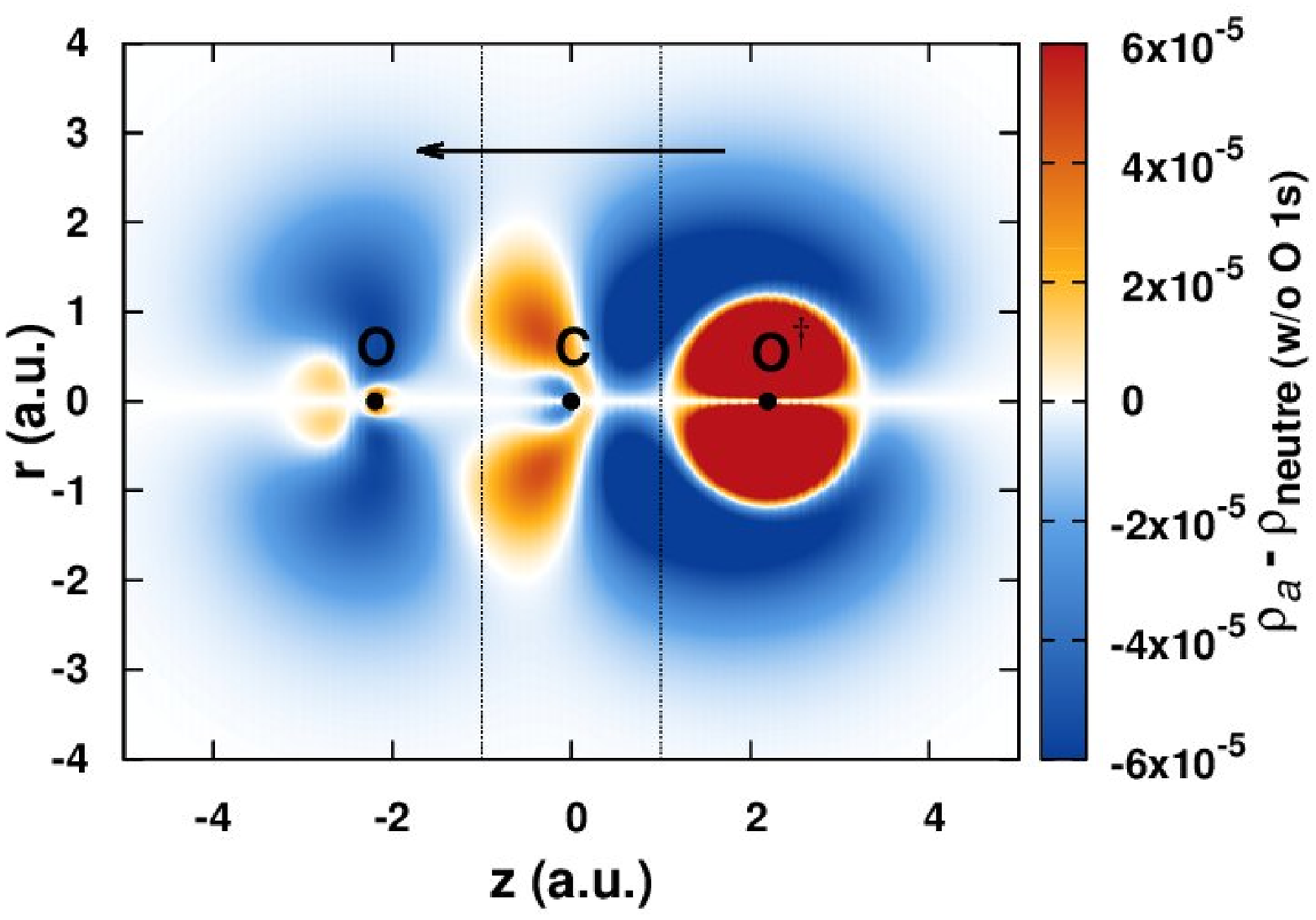}
        \caption{Cylindrically integrated difference in density between the unrelaxed \dch state (the initial neutral system without the two O 1s electrons) and the \dch ground state \textit{a}. The arrow length and direction are proportional to the variation of the dipole between the two considered states.}
       \label{relaxation_map}
\end{figure}

Regarding the satellites, as highlighted in our previous work \cite{FerPalPenIwaShiHikSoeItoLabTaiCar-JPCL-20}, the high intensity of the transition associated with a given shake-up state is the manifestation of the compensation between the shake-up induced electronic reorganization and the double core hole induced electronic relaxation.  Therefore, the large magnitude of the relaxation following the double core hole creation leads to the high relative intensity of the shake-up peaks with respect to the main line. We stress that the deviation from the sum rule on the transition amplitudes which can be seen in table \ref{table_CO2} and \ref{table_CO}, is a clear manifestation of the importance of the relaxation. In particular, important intensity transfer toward shake-off states were observed in the oxygen SCH-XPS spectra of \COO \cite{CarSelPen-PRA-16}. This is expected to be even more marked in the ssDCH case by the same reasoning as for the shake-up satellites. On the contrary, underestimating this deviation from the sum rule induces critical inaccuracy in the corresponding spectra~\cite{TasUedEha-CPL-11}.

We illustrate in Fig.\,\ref{relaxation_map} the double core hole induced electronic relaxation in \COO. To  this purpose, we report the difference in electron density (cylindrically integrated around the molecular axis) between the unrelaxed ssDCH system, which is the density of the neutral system minus the two O 1s electrons, and the fully relaxed \dch ground state (state~\textit{a}). The important positive (red) area around the rightmost oxygen illustrates the concentration of the electron density around the nucleus bearing the double core vacancy. Integration over all the positive (identically negative) regions provides a quantification of the double core hole induced relaxation. Through  this approach, we found that it engages a displacement of about~$2.2$~electrons, a magnitude similar to what was observed in the case of CO~\cite{FerPalPenIwaShiHikSoeItoLabTaiCar-JPCL-20}.

\begin{figure*}[t]
		\centering 
		 \includegraphics[scale=0.6]{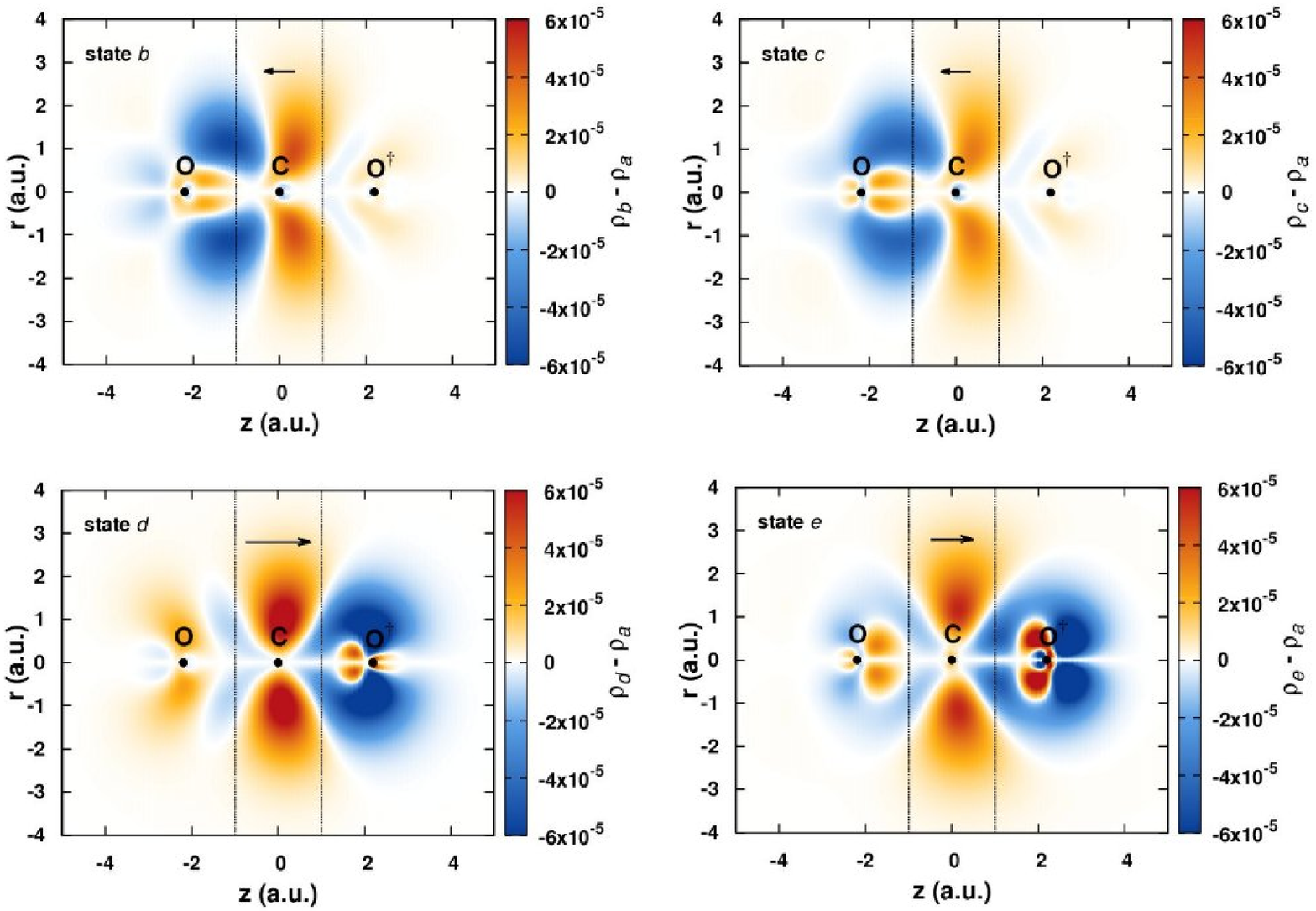}
        \caption{Cylindrically integrated difference in density between the \dch ground state \textit{a} and \textbf{top left\,)} shake-up state \textit{b}. \textbf{top right\,)}  shake-up state \textit{c}. \textbf{bottom left\,)}  shake-up state \textit{d}. \textbf{bottom right\,)} shake-up state \textit{e}. The arrows lengths and directions are proportional to the variation of the dipole between the two considered states.}
       \label{density_maps}
\end{figure*}

  Careful observation of Fig.\,\ref{relaxation_map} seems to indicate two different pathways through which the relaxation occurs. On one hand, the electron density strongly shrinks around the hollow center in response to the increased apparent nuclear charge. This contraction leads to the depletion of the electron cloud around the $\mathrm{O}^\dag$, including along the C$=$O$^\dag$ bond. On the other hand, a transfer of electron density from the leftmost oxygen atom toward the one bearing the double core vacancy takes place through the carbon atom  which acts here as a middleman. This last pathway is evidenced by the negative area surrounding the O nucleus and by the positive lobe on the C side of the O$=$C bond. 
  
  As the \dch ground state (\textit{a}) has an important single determinant character, most information regarding its electronic structure can be inferred from the relaxed MO reported in Fig.~\ref{mo_CO2}. As one can see, the $\pi_{\tsc{co}^\dag}$ orbitals are highly localized on the hollow O$^\dag$ center while the $\pi_{\tsc{oc}}$ orbitals are very well balanced along the whole O$=$C bond. Thanks to the relaxed structure of these MO, the electron density is efficiently shared between the carbon and the leftmost oxygen atoms while allowing for a major transfer toward the hollow O$^\dag$ atom. This unsurprising activation of the $\pi$ system upon electronic relaxation also yields consequences at the geometric level. As we discussed before, the geometric relaxation of the system following the double core ionization is predicted to lead to the desymmetrization of the two O$=$C and C$=$O$^\dag$ bonds. In particular we predicted a net contraction of the former while the later is expected to slightly expand (see table \ref{GS_geom}). Indeed, the very balanced $\pi_{\tsc{oc}}$ orbitals indicates a strong bond while the polarized, one could say slightly non-bonding, character of the $\pi_{\tsc{co}^\dag}$ indicates a weaker one. This illustrates the inducing effect of the electronic relaxation in regard to the geometrical relaxation.
  
  A refined quantification of the electronic reorganization was obtained by separating the entire space into three sections, as depicted by the thin vertical doted lines in Figs.\,\ref{relaxation_map} and \ref{density_maps}, and integrating over the positive and negative regions of the reported difference in electron density in each of these sections. We note that the choice to set the boundaries between the sections $1$~a.u. from the C nucleus on each side is arbitrary. It is only used to properly separate the different structures observed in Figs.\,\ref{relaxation_map} and \ref{density_maps}. The values of these integrals are reported in table \ref{table_integrals}.
  
 Following this spatial separation, the integral over the positive region in the rightmost section (thereafter dubbed as the $\mathrm{O}^\dag$ section) of Fig.\,\ref{relaxation_map} provides a good indicator of the amount of electron density that migrates closer to the hollow oxygen atom in order to compensate the release of two core electrons. We found for this integral a value of about $2.1$ electrons representing unsurprisingly the major part of the total double core hole induced relaxation. On one hand, the important negative integral in the central section indicates that a bit more than $0.4$ electron, initially close to the C nucleus (in particular along the C$=$O$^\dag$ bond) is relocated to compensate the double core vacancy on the O$^\dag$ atom. On the other hand, the negative integral in the leftmost section shows that about $0.7$ electron is transfered from the neutral O atom toward the hollow O$^\dag$ one. Finally, the minor electron density gain observed near the C atom and along the C$=$O bond was estimated to be about $0.1$ electron.
 
 To further illustrate the electron density relaxation following the double core ionization, we report the variation of the dipole induced by the relaxation. This is depicted in Fig.\,\ref{relaxation_map} by the arrow whose length and direction are proportional to the variation of the dipole between the fully relaxed \dch ``ground state" \textit{a}, and the unrelaxed ssDCH states, that is to says the initial neutral system in which we removed the two O core electrons. The direction of this arrow is therefore oriented in the opposite direction of the general electron motion and its length reflects the magnitude of the displacement. Numerical values of the dipoles and dipole variations for the different states discussed are reported in table \ref{table_dipole}. Here, the left orientation and the high amplitude of the dipole variation clearly depict the relaxation effects discussed before.

\begin{table*}[t]
\renewcommand{\arraystretch}{1.5}
\setlength{\tabcolsep}{0.2cm}
\caption{Integral of the negative and positive regions of the difference of density reported in Figs.\,\ref{relaxation_map} and \ref{density_maps}.}
\label{table_integrals}
\begin{tabular}{|c||cc|cc|cc|}
\hline
       & \multicolumn{2}{c|}{Leftmost O section}                                          & \multicolumn{2}{c|}{Central C section}                                           & \multicolumn{2}{c|}{Rightmost $\mathrm{O}^\dag$ section}                                         \\
       & \multicolumn{1}{l}{Negative int.} & \multicolumn{1}{l|}{Positive int.} & \multicolumn{1}{l}{Negative int.}     & \multicolumn{1}{l|}{Positive int.} & \multicolumn{1}{l}{Negative int.}     & \multicolumn{1}{l|}{Positive int.} \\ \hline \hline
Fig.\,\ref{relaxation_map} & \multicolumn{1}{c}{$-0.685$}                & $0.011$                                      & \multicolumn{1}{c}{$-0.447$}                & $0.123$                                      & \multicolumn{1}{c}{$-1.043$}                & $2.072$                                      \\ 
\shortstack{\vspace{+0.1cm}\\Fig.\,\ref{density_maps}\\~\\Top Left (\textit{b})} & \multicolumn{1}{c}{$-0.288$}                & $0.040$                                      & \multicolumn{1}{c}{$-0.128$}                & $0.280$                                      & \multicolumn{1}{c}{$-0.004$}                & $0.081$                                      \\ 
Top Right (\textit{c}) & \multicolumn{1}{c}{$-0.289$}                & $0.043$                                      & \multicolumn{1}{c}{$-0.107$}                & $0.237$                                      & \multicolumn{1}{c}{$-0.005$}                & $0.077$                                      \\ 
Bottom Left (\textit{d}) & \multicolumn{1}{c}{$-0.018$}                & $0.146$                                      & \multicolumn{1}{c}{$-0.030$}                & $0.493$                                      & \multicolumn{1}{c}{$-0.608$}                & $0.040$                                      \\ 
Bottom Right (\textit{e}) & \multicolumn{1}{c}{$-0.072$}                & $0.064$                                      & \multicolumn{1}{c}{$-0.061$}                & $0.406$                                      & \multicolumn{1}{c}{$-0.558$}                & $0.136$                                      \\ \hline
\end{tabular}
\end{table*}

\subsubsection{Nature of the main shake-up satellites}
\label{CO2_satellites}

According to our interpretation of the monopolar form of the transition amplitude \eqref{overlap}, the shake-up states associated with the important peaks in the ssDCH spectrum should be the one associated with a reorganization of the electron density which counteracts the previously described double core hole induced relaxation effects. This is because such shake-up induced reorganization of the electronic structure will allow the excited \dch state to revert towards a structure similar to the one characterizing the initial neutral system. Therefore, even though the ground state \dch remains the most probable final state to be formed because it is still very similar to the neutral ground state, the important relaxation induced by the formation of the double core vacancy allows for the important relative intensity of the shake-up satellites in a \mbox{ssDCH} spectrum.

Unsurprisingly, the shake-up states \textit{b} and \textit{c}, apart from being close in energy, are also very similar in the type of electronic reorganization that characterizes them. From table \ref{table_CO2}, one can see that both wave functions describing these states, (Eq.\,\ref{K-2_wf}), are strongly dominated by Slater determinants (in the relaxed ssDCH optimized MO basis \{$v$\}) corresponding to single and double excitation from the occupied {$2\,\pi_{\tsc{oc}}$, $2'\pi_{\tsc{oc}}$} and $4\,\sigma_{\tsc{oc}}^*$ molecular orbitals toward the virtual {$3\,{\pi^*}$, $3'{\pi^*}$} and $5\,{\sigma}$ MOs. We also note the important participation of the ``HF \dch determinant" (the \dch determinant in the relaxed basis \{$v$\} with the lowest energy) in the CI expansion of state \textit{b}. Such determinants characterized by excitation from molecular orbitals localized on the O and C atoms towards MOs globally delocalized over the whole molecular system (with a predominant C character for the anti-bonding $\pi^*$ ones) indicate a transfer of electron density from the O$=$C bond region toward C$=$O$^\dag$ one.

\begin{table}[t]
\begin{threeparttable}

\renewcommand{\arraystretch}{1.35}
\setlength{\tabcolsep}{0.15cm}
\caption{Projection of the dipole on the z axis for the states associated with a strong line in the O \dch spectrum of \COO and of the unrelaxed ssDCH system.}

\begin{tabular}{|c||cc|}
\hline
Reference state & Dipole\tnote{(a)}\ \ \ (D) & Dipole variation\tnote{(b)}\ \ \ (D) \\
\hline
\hline
Unrelaxed \dch   & 11.142 & - \\
State \textit{a} &  2.408 & $-$8.735\\
State \textit{b} &  0.589 & $-$1.818\\
State \textit{c} &  0.672 & $-$1.735\\
State \textit{d} &  6.219 & $+$3.811\\
State \textit{e} &  4.865 & $+$2.457\\
\hline
\end{tabular}

\begin{tablenotes}
	\item[a] Origin of the space taken on the C nucleus.
	\item[b] With respect to the unrelaxed \dch system for state~\textit{a} and to state \textit{a} for all the shake-up states (\textit{b} to \textit{e}).
\end{tablenotes}

\label{table_dipole}

\end{threeparttable}
\end{table}

This electron density transfer can be observed in the two top panels of Fig.\,\ref{density_maps} which illustrate the shake-up induced electronic reorganization by showing the difference in electron density between the excited shake-up state considered and the fully relaxed \dch ground state \textit{a}. Integrals over the complete space indicate that the shake-up states \textit{b} and \textit{c} are both characterized by a reorganization of the density accounting for about $0.4$ electron with respect to the \dch ground state \textit{a}. Comparison between Figs.\,\ref{relaxation_map} and \ref{density_maps} and inspection of the section integrals reported in table~\ref{table_integrals} show that in both cases the electronic reorganization induces a partial compensation of the relaxation of the electron density in the vicinity of the C atom. Indeed, one can see that the positive region in the central section (C section) of the two top panels of Fig.\,\ref{density_maps},  whose integral represents respectively about $0.2$ and $0.3$ electron, compensates for more than half of the negative region on the right side of the ``C section" of Fig.\,\ref{relaxation_map}. In addition, the negative regions of the same section, respectively representing a displacement of about $0.1$ electron for both shake-up states, almost perfectly counteract the positive lobe in the vicinity of the C observed in the double core hole induced relaxation. The dipole variations induced by the shake up reorganization, obtained as the difference between the dipole of the concerned shake-up states and of the fully relaxed \dch  state \textit{a} and represented by the arrows in the respective panels of Fig.\,\ref{density_maps}, support these interpretations.

Apart from the compensation of the loss in electron density in the vicinity of the C nucleus, the electron density transfer also tends to restore the broken symmetry of the \COO molecule. Indeed, one can estimate the total electron density variation with respect to the neutral system in one section of the space and for a given final shake-up state by adding the effects of the relaxation, depicted in figure~\ref{relaxation_map}, with the shake-up specific electron density reorganization of figure~\ref{density_maps}. By doing so, one can see that the total variation of electron density with respect to the initial neutral system in the left section (O section) of the space  is about $-1.0$ electron for state \textit{b} and $-0.9$ for state \textit{c}. Similarly, one can evaluate the total electron density variation with respect to the neutral system in the $\mathrm{O}^\dag$ section and find a difference of about $-0.9$ for both states  \textit{b} and \textit{c}. We note that for evaluating the total density variation, we considered that the 2 electron loss due to the formation of the double core vacancies was fully localized in the $\mathrm{O}^\dag$ section. Therefore, one can see that, when it comes to the raw amount of electron, the density characterizing the shake-up states \textit{b} and \textit{c} is close to be well symmetrically distributed between the two oxygen atoms. The dipoles reported in table \ref{table_dipole} depict an analogue picture. Indeed, one can see that both states \textit{b} and \textit{c} are characterized by values much smaller than for state \textit{a} showing that the electron density is much more symmetrically distributed in the two discussed shake-up states than in the fully relaxed ssDCH.

The two shake-up states \textit{d} and \textit{e} are also akin to each other. Both of the wave functions characterizing these states are strongly dominated by single excitations from orbitals with a strong C and O$^\dag$ character, in particular the {$1\,\pi_{\tsc{co\tsp{\dag}}}$, $1'\pi_{\tsc{co\tsp{\dag}}}$} and $3\,\sigma_{\tsc{co\tsp{\dag}}}^*$ orbitals, toward the same three delocalized ones as discussed before ({$3\,{\pi^*}$, $3'{\pi^*}$} and $5\,{\sigma}$). Such determinants indicate therefore the symmetrical effect of the one previously discussed, that is a displacement of electron density from the C$=$O$^\dag$ bond region towards the O$=$C one.

Once again, the shake-up induced reorganization of the electron density associated with state \textit{d} and \textit{e}, depicted in the two bottom panels of Fig.\,\ref{density_maps}, matches with the interpretation produced by looking at the main determinants in their wave function. For both states, this reorganization is shown to involve a displacement of about $0.7$ electron. The most notable effect in comparison to the double core hole induced relaxation is the partial compensation of the concentration of electron density around the hollow O$^\dag$. Indeed, table \ref{table_integrals} shows that the total variation of electron density in the ``$\mathrm{O}^\dag$ section" of these two panels (negative integral + positive integral) amounts to a loss of about $0.6$ electron for state \textit{d} and $0.4$ electron for state \textit{e} which more or less represent half of the relaxation induced electron density gain in the same section of Fig.\,\ref{relaxation_map} ($\sim 1.0$ electron). In conjunction, the total shake-up induced gain of electron density depicted in the ``C sections" of the two bottom panels of Fig\,\ref{density_maps}, about $0.5$ and $0.3$ electron respectively for state \textit{d} and \textit{e}, totally compensates the loss in electron density in the vicinity of the C atom following the complete double core hole induced relaxation ($\sim 0.3$ electron).

\begin{table*}[t]
\begin{threeparttable}
\renewcommand{\arraystretch}{1.9}
\setlength{\tabcolsep}{0.25cm}
\caption{NOTA+CIPSI transition amplitudes and relative binding energies associated with the intense peaks of the O \dch spectrum of CO, and their corresponding final ssDCH state. Reproduced from ref.\,\onlinecite{FerPalPenIwaShiHikSoeItoLabTaiCar-JPCL-20}.}
\begin{tabular}{| c c c p{10.5 cm} |}
	\hline	
    Label\tnote{(a)} & (BE$-$DCIP) & $|T_{\textsc{i} \rightarrow \textsc{f},\mathrm{n}}|^2$ & Main configurations\tnote{(b,c)} \\
    \hline \hline
    \textit{a\,$'$} & 0      & 0.205 &  $+$ 0.94\,\big\{$(\mathrm{1s\tsb{\textsc{o}}})^{0}$\,\big\} \\
    \textit{b\,$'$} & 15.214 & 0.010 &  $-$ 0.63\,\big\{$(\mathrm{1s\tsb{\textsc{o}}})^{0}$ $(3\sigma)^{0}$ $(2\,\pi^*)^{2}$\big\} $-$ 0.63\,\big\{$(\mathrm{1s\tsb{\textsc{o}}})^{0}$ $(3\sigma)^{0}$ $(2'\pi^*)^{2}$\big\} $-$ 0.22\,\big\{$(\mathrm{1s\tsb{\textsc{o}}})^{0}$\big\}\\
    \textit{c\,$'$} & 22.631 & 0.132 & \baselineskip=15pt $-$ 0.35\,\big\{$(\mathrm{1s\tsb{\textsc{o}}})^{0}$ $(1\,\pi)^{1}$ $(2\,\pi^*)^{1}$\,\big\} $-$ 0.35\,\big\{$(\mathrm{1s\tsb{\textsc{o}}})^{0}$ $(1'\pi)^{1}$ $(2'\pi^*)^{1}$\,\big\} \hphantom{-----------.-----} $+$ 0.29\,\big\{$(\mathrm{1s\tsb{\textsc{o}}})^{0}$ $(3\,\sigma)^{1}$ $(5\,\sigma^*)^{1}$\,\big\} $-$ 0.14\,\big\{$(\mathrm{1s\tsb{\textsc{o}}})^{0}$ $(1\,\pi)^{1}$ $(2'\pi^*)^{1}$\,\big\} \hphantom{----------------} $+$ 0.14\,\big\{$(\mathrm{1s\tsb{\textsc{o}}})^{0}$ $(1'\pi)^{1}$ $(2\,\pi^*)^{1}$\big\} \hphantom{---------------------------------..---------.-.----------} \\
    \textit{d\,$'$} & 37.582 & 0.025 & \baselineskip=15pt $-$ 0.34\,\big\{$(\mathrm{1s\tsb{\textsc{o}}})^{0}$ $(1\,\pi)^{1}$ $(3'\pi)^{1}$\ \big\} $+$ 0.34\,\big\{$(\mathrm{1s\tsb{\textsc{o}}})^{0}$ $(1'\pi)^{1}$ $(3\,\pi)^{1}$\,\big\} \hphantom{----------------} $-$~0.19\,\big\{$(\mathrm{1s\tsb{\textsc{o}}})^{0}$ $(2\,\sigma)^{1}$ $(5\,\sigma^*)^{1}$\big\} $-$ 0.14\,\big\{$(\mathrm{1s\tsb{\textsc{o}}})^{0}$ $(1\,\pi)^{1}$ $(3\,\pi)^{1}$\,\big\} \hphantom{----------------} $-$~0.14\,\big\{$(\mathrm{1s\tsb{\textsc{o}}})^{0}$ $(1'\pi)^{1}$ $(3'\pi)^{1}$\hphantom{..}\big\}\,\,\,$+$ 0.11\,\big\{$(\mathrm{1s\tsb{\textsc{o}}})^{0}$ $(2\sigma)^{1}$ $(7\sigma^*)^{1}$\,\big\}\\
    \textit{e\,$'$} & 40.579 & 0.012 & $-$ 0.55\,\big\{$(\mathrm{1s\tsb{\textsc{o}}})^{0}$ $(2\sigma)^{1}$ $(6\sigma^*)^{1}$\,\big\} $-$ 0.19\,\big\{$(\mathrm{1s\tsb{\textsc{o}}})^{0}$ $(2\sigma)^{1}$ $(5\sigma^*)^{1}$\,\big\} \hphantom{------------.--------} \\
    \hline
\end{tabular}

\begin{tablenotes}
	\item[a] Only states with $|T_{\textsc{i} \rightarrow \textsc{f},\mathrm{n}}|^2 > 0.01$
	\item[b] Only configurations with CI coefficients in intermediate normalization $C_\nu^n / \mathrm{max}\{C^n_\nu\} > 0.3 $
	\item[c] Open shell configurations are only reported once but appear twice in the expansion because of the $\alpha$ $\beta$ symmetry. 
\end{tablenotes}

\label{table_CO}

\end{threeparttable}
\end{table*}

Contrary to the two previous shake-up states \textit{b} and \textit{c}, the shake-up induced electron density reorganization observed in the case of states \textit{d} and \textit{e} tends to amplify the symmetry breaking between the two O atoms. For these two states, the total electron density variation with respect to the neutral system in the ``O section" of the space can be estimated to represent a loss of about $0.5$ and $0.7$ electron respectively, while the loss in electron density in the ``$\mathrm{O}^\dag$ section" is evaluated to be about $1.5$ electron for state \textit{d} and $1.4$ electron for state \textit{e} yielding to a raw difference in electron density  around the two oxygen nucleus close to be equivalent to 1 electron. Similarly to the previous discussion regarding the shake-up states \textit{b} and \textit{c}, the shake-up induced electronic reorganization effects are well captured by the variation of the dipole and the increase in magnitude for states \textit{d} and \textit{e} with respect to state \textit{a} reflects the increase in the unbalanced of the electron density repartition between the two oxygen atoms.

In summary, the two main shake-up satellites structures observed in the O \dch spectrum of \COO are the signature of two distinctive type of electron density reorganization induced by the shake-up excitations. The first one is mainly characterized by a transfer of electron density from the leftmost core-filled O atom toward the C atom while the second one correspondse to the retrocession of the electron density from the hollow O$^\dag$ atom toward the C atom.

\subsection{Comparison between the O \dch spectra of CO\tsb2 and CO}
\label{CO_vs_CO2}

\begin{figure*}[t]
        \centering
        \includegraphics[scale=0.40]{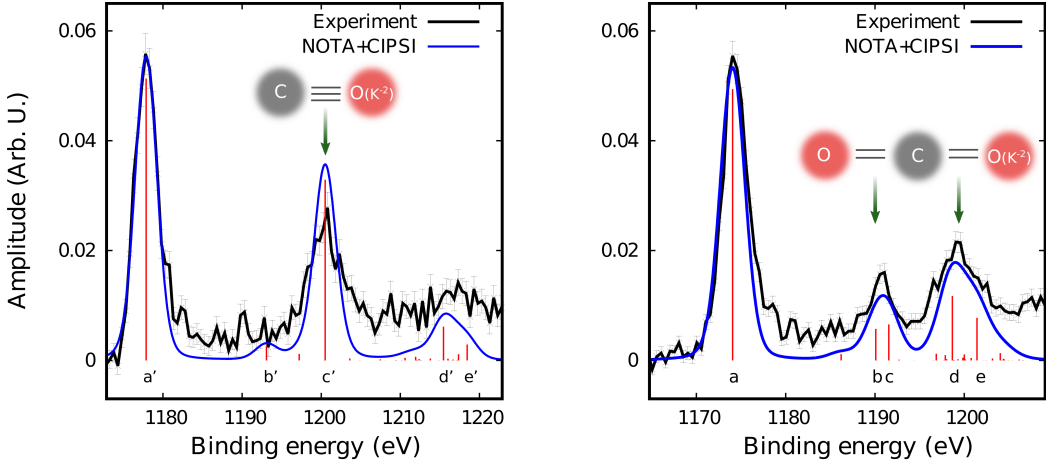} 
        \caption{Comparison between the O~\dch spectrum of CO (left) and \COO (right). NOTA+CIPSI spectra (blue), experimental spectra (black) and $0.25 \times |T_{\textsc{i} \rightarrow \textsc{f},n}|^2$ (red vertical lines) are represented for both systems.}
        \label{plot_CO_vs_CO2}
\end{figure*}

The O ssDCH spectrum of the CO\tsb{2} molecule is characterized by a more complex and richer satellite structure than the O \dch spectrum of CO that we presented in our previous paper \cite{FerPalPenIwaShiHikSoeItoLabTaiCar-JPCL-20}. Nevertheless the two spectra still hold some similarities. We report for comparison the NOTA+CIPSI O \dch spectrum of CO in Fig.\,\ref{plot_CO_vs_CO2} and table~\ref{table_CO}.

Because the approximation of Eq.\,\ref{overlap} relies on considering constant the two electron dipole term which should be very system dependent, our method only yields transition intensities  which are comparable within a given ssDCH spectrum. Therefore, we stress that a direct comparison of the reported intensity of the CO and \COO spectra is not possible. Still, relative peaks intensity of both spectra taken independently are trustworthy. Identically, the  reported absolute DCIP as well as the difference of DCIP between CO and \COO are also reliable as shown in section \ref{new_section_DCIP}.

Both spectra display a similar intense satellite region around 1200~eV. In contrast with the CO spectrum where this satellite region was consecutive to the transition to only one specific cation state (\textit{c}\,$'$), two shake-up states (\textit{d} and \textit{e}) are associated with the main transitions responsible for this satellite region in the CO\tsb2 spectrum. However the CO spectrum does not display an equivalent to the intense satellite region located around 1190~eV in the CO\tsb2 spectrum.

From inspection of table \ref{table_CO2} and \ref{table_CO}, one can see that the \COO states \textit{d} and \textit{e} share with the CO \dch shake-up \textit{c}\,$'$ more than their binding energies. Indeed, both \textit{d} and \textit{e} states wave functions are dominated by single excitation from orbitals with a strong O$^\dag$ character to orbitals mainly localized around the C, preferably from $\pi_{\tsc{c}\tsc{o}^\dag}$~to~$\pi^*$ and $\sigma_{\tsc{o}^\dag\tsc{c}}$~to~$\sigma^*$. Similarly, the CO core dication state \textit{c}\,$'$ is also strongly  dominated by the same kind of excited determinants. Thus, one can see that these two satellite regions both arise from transition to the same type of shake-up states characterized by a transfer of electron density from the hollow $\mathrm{O}^\dag$ atom towards its C neighbor.   

The first \COO satellite region (stemming from the transition to \textit{b} and \textit{c}) do not have an obvious correspondence in the CO spectrum. One could be tempted to connect it with the small shake-up \textit{b}\,$'$. However, even if one notices a similar structure of the CI expansion of the wave functions describing state \textit{b} and \textit{b}\,$'$ (important CI coefficients on double excitation and a significant weight of the ``HF \dch determinant") no further similarity exists between these two satellite regions. In fact, as it was said before, \textit{b} and \textit{c} are mainly characterized by an electron density transfer from the core filled O atom toward the C. Obviously this type of electron density reorganization observed in \COO does not have any counterpart in the CO molecule.

Here, we showed that in \COO, the second O atom can act as an electron reservoir which allows for more than one way for the electron density to reorganize. The added flexibility of the system, in comparison to the CO molecule, leads to a richer satellite structure. Therefore, one can interpret the two main shake-up satellite regions in the O \dch spectrum of \COO as the signature of the two non equivalent O$=$C and C$=$O$^\dag$ bonds.

\section{Conclusion}
\label{conclusion}
 In this work we used the recently introduced NOTA+CIPSI method to investigate the signature of the electronic relaxation and reorganization in the O ssDCH shake-up spectrum of \COO. 
 
Similarly~to what we already observed for CO \cite{FerPalPenIwaShiHikSoeItoLabTaiCar-JPCL-20}, we found that the NOTA+CIPSI spectrum precisely reproduces the experimental spectrum of \COO. This consolidates the status of the NOTA+CIPSI as a high accuracy computation method able to yield exact peak intensity thanks to the use of two non mutually orthogonal MO basis sets, and peak position thanks to a well balanced description of initial neutral state and final ssDCH states. On this topic, we produced clear illustration of the sweet converging behavior of the peak position in the NOTA+CIPSI approach as well as the important speed-up granted by the perturbative correction on the energy.
 
The high quality descriptions of the wave functions within the NOTA+CIPSI method allows for in-depth investigation regarding the produced states. We successfully applied to the O \dch satellites of \COO our interpretation of the high intensity transition associated with specific shake-up state as the consequence of the compensation between the double core hole induced relaxation and the shake-up induced electronic reorganization further bolstering it as a general behavior. We highlighted the striking similarities as well as the interesting differences between the two main shake-up satellite regions observed in the O \dch spectrum of \COO with the one main satellite of the CO spectrum. In particular, we showed that the two main \COO satellite regions correspond to the two non equivalent oxygen-carbon bonds in the ssDCH \COO. This suggests that the satellite structures observed via ssDCH spectroscopy, thanks to the magnification of the shake-up intensity, might be a good indicator of the bond structure of the system studied.
 
\section*{Conflicts of interest}
There are no conflicts of any sort to declare.

\begin{acknowledgements}
We are very grateful to Kenji Ito for its participation to the experimental measurements.

We thanks Labex MiChem, part of French state funds managed by the ANR within the Programme d'Investissements d'Avenir (Sorbonne Universit\'e, ANR-11-IDEX-0004-02), that provided PhD funding for \mbox{A. Fert\'e}. 

K. Soejima acknowledges the support of the Labex Plas@Par, managed by the Agence Nationale de la Recherche as part of the Programme d'Investissements d'Avenir under Reference ANR-11-IDEX-0004-02. 

The authors are appreciative of the Photon Factory staff for the stable operation of the storage ring. This work was performed with the approval of the PF Program Advisory Committee (Proposal 2010G621).

\end{acknowledgements}



\end{document}